\documentclass[draftclsnofoot]{IEEEtran}
\onecolumn
\usepackage{amsmath, amssymb}
\usepackage{amsthm}
\usepackage{cite}
\usepackage{bm}
\usepackage[dvips]{graphics}
\usepackage{graphicx}
\usepackage{psfrag}

\newtheorem{theorem}{Theorem}
\newtheorem{lemma}{Lemma}

\newtheorem{proposition}{Proposition}
\theoremstyle{definition}
\newtheorem{definition}{Definition}
\newtheorem*{remarks}{Remarks}
\newtheorem{example}{Example}

\newcommand{\eq}[1]{(\ref{#1})}

\newcommand{\eps}{\varepsilon}
\newcommand{\suchthat}{\colon}
\newcommand{\tv}{\mathbf{t}}
\newcommand{\Tc}{\mathcal{T}}

\begin{document}
\allowdisplaybreaks
\title{Directed Information, Causal Estimation, and
Communication in Continuous Time}

\author{Tsachy Weissman, Young-Han Kim and Haim H.~Permuter
\thanks{This work is partially supported by the NSF grant CCF-0729195,
BSF grant 2008402, and the Center for Science of Information (CSoI),
an NSF Science and Technology Center, under grant agreement CCF-0939370.
H.~H.~Permuter has been partially supported by the Marie Curie Reintegration
fellowship. Author's emails: tsachy@stanford.edu, yhk@ucsd.edu, haimp@bgu.ac.il}}

\maketitle

\begin{abstract}
A notion of directed information between two continuous-time processes
is proposed. A key component in the definition is taking an infimum over
all possible partitions of the time interval, which plays a role no less
significant than the supremum over ``space'' partitions inherent in the
definition of mutual information.
Properties and operational interpretations in estimation and communication
are then established for the proposed notion of directed information.
For the continuous-time additive white Gaussian noise channel,
it is shown that Duncan's  classical relationship between
causal estimation error and mutual information continues
to hold in the presence of
feedback upon replacing mutual information by directed information. A
parallel result is established
for the Poisson channel.
The utility of this relationship is  demonstrated in computing
the directed information rate between the input and output processes
of a continuous-time
Poisson channel with feedback, where the channel input process is constrained to be
constant between events at the channel output.
Finally, the capacity of a wide class of continuous-time channels with feedback
is established via directed information, characterizing
the fundamental limit on reliable communication.
\end{abstract}

\begin{keywords}
Causal estimation, conditional mutual information, continuous time, directed information, Duncan's theorem,
feedback capacity, Gaussian channel, Poisson channel, time partition.
\end{keywords}

\section{Introduction}
The directed information $I(X^n \to Y^n)$ between two random $n$-sequences
$X^n = (X_1,\ldots, X_n)$ and $Y^n = (Y_1,\ldots,Y_n)$ is a natural
generalization of Shannon's mutual information to random objects obeying
causal relations. Introduced by Massey~\cite{Massey90}, this notion
 has been shown to arise as the canonical
answer to a variety of problems with causally dependent components.
For example, it plays a pivotal role in characterizing the capacity
$C_\text{FB}$ of a communication channel with
feedback. Massey~\cite{Massey90} showed that the feedback capacity
is upper bounded as
\begin{equation} \label{eq:massey_ub}
C_\text{FB} \le \lim_{n\to\infty} \max_{p(x^n||y^{n-1})}\frac{1}{n}
I(X^n \to Y^n),
\end{equation}
where $I(X^n\to Y^n) = \sum_{i=1}^n I(X^i; Y_i|Y^{i-1})$ and
$p(x^n||y^{n-1}) = \prod_{i=1}^{n} p(x_i|x^{i-1},y^{i-1})$;
see also Kramer~\cite{Kramer03} that streamlines the
notion of directed information by causal conditioning.
The upper bound in~\eqref{eq:massey_ub} is tight
for certain classes of ergodic channels, such as
general nonanticipatory channels satisfying certain
regularity conditions~\cite{TatikondaMitter_IT09},
channels with finite input memory and ergodic noise~\cite{Kim07_feedback},
and indecomposable finite-state channels~\cite{PermuterWeissmanGoldsmith09},
paving the road to a computable characterization of feedback capacity;
see~\cite{Chen05, yhk06, PermuterCuffVanRoyWeissman08} for examples.

Directed information and its variants also characterize (via multiletter expressions) the
capacity for two-way channels~\cite{Kramer03},
multiple access channels with feedback~\cite{Kramer03,
PermuterWeissmanChenMAC_IT09}, broadcast channels with feedback
\cite{DaborahGoldsmith10BC_feedback_really}, and compound channels with feedback
\cite{ShraderPermuter09CompoundIT}, as well as the rate--distortion function with feedforward
\cite{PradhanVenkataramananIT_feedforward07,PradhanIT_Gaussianfeedforward07}. In another context,
directed information captures the difference in growth rates of wealth in horse race
gambling due to \emph{causal} side information~\cite{PermuterKimTsachy08ISIT}. This provides a
natural interpretation of $I(X^n \to Y^n)$ as the amount of information about $Y^n$
causally provided by $X^n$ on the fly.  Similar interpretations for directed information can be drawn for other problems in science and engineering~\cite{KPW08}.

This paper is dedicated to extending the mathematical notion of directed information
to continuous-time random processes and to establishing results
that demonstrate the operational significance of this notion in
estimation and communication.
Our contributions include the following:
\begin{itemize}
\item
We introduce the notion of directed information in continuous time.
Given a pair of continuous-time processes in a time interval and its partition consisting
of $n$ subintervals,  we first consider the (discrete-time) directed information
for the two sequences of length $n$ whose components are the sample paths on the respective subintervals.
The resulting quantity depends on the specific partition of the time interval.
We define directed information in continuous time by taking the infimum over all
finite time partitions.
Thus, in contrast to mutual information in continuous time which can be
defined as a \emph{supremum} of mutual information over finite ``space''
partitions~\cite[Ch.~2.5]{Gallager68}, \cite[Ch.~3.5]{Pinsker60},
inherent to our notion of directed information is a similar supremum
followed by an \emph{infimum} over time partitions.
We explain why this definition is natural by showing that the continuous-time directed information
inherits key properties of its discrete-time origin and by establishing
new properties that are meaningful in continuous time.

\item We show that this notion of directed information arises in extending
classical relationships between information and estimation in
continuous time---Duncan's theorem \cite{Duncan1968} that
relates the minimum mean squared error (MMSE) in causal estimation of
a target signal based on an observation through an additive
white Gaussian noise channel to the \emph{mutual information} between the target signal and the observation,
and its counterpart for the Poisson channel---to the scenarios in which the channel
input process can causally depend on the channel output process, whereby
corresponding relationships now hold between \emph{directed information}
and estimation.

\item We illustrate these relationships between directed information and estimation
by characterizing the directed information rate and the feedback capacity of a continuous-time
Poisson channel with inputs constrained to be constant between events at the channel output.

\item We establish the fundamental role of continuous-time
directed information in characterizing the feedback capacity of a large class of continuous-time
channels. In particular, we show that for channels where the output is a function of the input and
some stationary ergodic ``noise'' process, the continuous-time directed information characterizes the
feedback capacity of the channel.
\end{itemize}

The remainder of the paper is organized as follows. Section \ref{sec: Definition of Directed
Information in Continuous Time} is devoted to the definition of directed information and related
quantities in continuous time, which is followed by a presentation of key properties of
continuous-time directed information in Section \ref{sec: Properties of the Directed Information
in Continuous Time}. In Section \ref{sec: Directed Information and Causal Estimation}, we
establish the generalizations of Duncan's theorem and its Poisson counterpart that accommodate the presence of feedback.
In Section \ref{sec: poisson feedback example}, we apply the relationship
between the causal estimation error and directed information for the Poisson channel
to compute the directed information rate between the input and the output of this channel in a scenario
that involves feedback.
In Section \ref{sec: Communication Over Continuous-Time Channels with Feedback}, we study a general
feedback communication problem in which our notion of directed information in continuous time emerges naturally in the characterization of the feedback capacity.
Section \ref{sec: Concluding Remarks} concludes the paper with a few remarks.


\section{Definition and Representation of Directed Information in Continuous Time}
\label{sec: Definition of Directed Information in Continuous Time}

Let $P$ and $Q$ be two probability measures on the same space
and $\frac{d P}{d Q}$ be the Radon--Nikodym derivative of $P$ with respect to $Q$.
The relative entropy between $P$ and $Q$ is defined as
\begin{equation}\label{eq: relative entropy}
D (P \| Q )
:=  \begin{cases}
 \int \bigl( \log \frac{d P}{d Q} \bigr) \, d P &   \mbox{if $\frac{d P}{d Q}$ exists,}    \\
  \infty &  \mbox{otherwise.}
\end{cases}
\end{equation}
For jointly distributed random objects $U$ and $V$, the mutual information between them is defined as
\begin{equation}\label{eq: mutinf defined}
I(U; V) := D (P_{U,V} \| P_U \times P_V ),
\end{equation}
where $P_U \times P_V$ denotes the product distribution under which $U$ and $V$ are
independent but maintain their respective marginal distributions.
As an alternative, the mutual information is defined~\cite[Ch.~2.5]{Gallager68} as
\begin{equation}\label{eq: mutinf defined ALT}
I(U; V) := \sup I([U];[V]),
\end{equation}
where the supremum is over all finite quantizations of $U$ and
$V$. That the two notions coincide has been established in, e.g.,
\cite{Kolmogorov1956}, \cite[Ch.~3.5]{Pinsker60}.
We write $I(P_{U,V})$ instead of $I(U; V)$ when we wish to emphasize
the dependence on the joint distribution $P_{U,V}$.

For a jointly distributed random triple $(U,V,W)$
with components in arbitrary measurable spaces,
we define
the conditional mutual information between $U$ and $V$ given $W$
as
\begin{equation}\label{eq: conditional mutinf defined}
I(U; V|W) := \sup I([U];[V] | W)  ,
\end{equation}
where the supremum is over all finite quantizations of $U$ and $V$. This quantity, due to Wyner~\cite{Wyner78_def_of_conditional_mutual}, is always well defined and
satisfies all the basic properties of conditional mutual information for discrete and continuous random variables,
in particular:
\begin{enumerate}
\item \emph{Nonnegativity:} $I(U; V | W) \ge 0$ with equality iff $U \to W \to V$ form a Markov chain (that is,
$U$ and $V$ are conditionally independent given $W$).

\item \emph{Chain rule:} $I(U; V, X | W) = I(U; V | W) + I(U; X | V, W)$.

\item \emph{Data processing inequality:} If $U \to (W,X) \to V$ form a Markov chain, then $I(U; X | W) \ge I(U; V | W)$
with equality iff $I(U; V |  W, X) = 0$.
\end{enumerate}
The definition in \eqref{eq: conditional mutinf defined} coincides with
Dobrushin's more restrictive
definition \cite[p.~29]{Pinsker60}
\begin{equation}
\int I(P_{U,V|W=w}) \, d P_W (w),
\end{equation}
where $P_{U,V|W=w}$ is a regular version of the conditional probability law
of $(U,V)$ given $\{W = w\}$ (cf.~\cite[Ch.~6]{Kallenberg2002}) if it exists.


Let $(X^n, Y^n)$ be a pair of random $n$-sequences. The directed
information from $X^n$ to $Y^n$ is defined as
\begin{equation} \label{eq: directed info in discrete time defined}
I(X^n \to Y^n) := \sum_{i=1}^n I(X^i;Y_i|Y^{i-1}).
\end{equation}
Note that, unlike mutual information, directed information is
asymmetric in its arguments, i.e., $I(X^n \to Y^n) \ne I(Y^n \to X^n)$ in general.

Let us now develop the notion of directed information between two continuous-time stochastic
processes on the time interval $[0,T)$.  For a continuous-time process $\{ X_t \}$, let
$X_a^b  = \{ X_s \suchthat a \le s < b \}$
denote the process in the time interval $[a, b)$.
Let $\tv = (t_0, t_1, \ldots, t_n )$ denote a
vector with components satisfying
\begin{equation}\label{eq: what components of t satisfy}
    0 = t_0 < t_1 < \cdots < t_n = T.
\end{equation}
Let $X_0^{T, \tv}$
denote the sequence of length $n$
resulting from ``chopping up'' the continuous-time signal $X_0^T$
into consecutive segments as
\begin{equation}\label{eq: eps chopped up signal}
X_0^{T, \tv}
= ( X_0^{t_1}, X_{t_1}^{t_2}, \ldots , X_{t_{n-1}}^T).
\end{equation}
Note that each component of the sequence is a continuous-time stochastic
process.  For a pair of jointly distributed stochastic processes $(X_0^{T} , Y_0^{T} )$,  define
\begin{align}
I_{\tv} ( X_0^T \to Y_0^T )
&:= I ( X_0^{T, \tv} \to Y_0^{T, \tv} )
    \label{eq: directed info between sequences} \\
&= \sum_{i=1}^{n} I ( Y_{t_{i-1}}^{t_i}; X_0^{t_i} \big| Y_0^{t_{i-1}}),
    \label{eq: mut inf summands in definition of ieps}
\end{align}
where on the right side of \eqref{eq: directed info between sequences} is the directed
information between two sequences of length $n$ defined
in~\eqref{eq: directed info in discrete time defined};
and in \eqref{eq: mut inf summands in definition of ieps}
we note that the conditional mutual information terms, defined as in  (\ref{eq: conditional mutinf defined}), are between
two continuous-time processes, conditioned on a third.
We extend this definition to
$I _{\tv} ( X_0^T \to Y_0^T | V)$, where $V$ is a random object jointly distributed with $(X_0^T, Y_0^T)$, in the obvious way, namely
\begin{align}
I_{\tv} ( X_0^T \to Y_0^T |V )
&:= I ( X_0^{T, \tv} \to Y_0^{T, \tv} |V )
    \label{eq: directed info between sequences} \\
&:= \sum_{i=1}^{n} I ( Y_{t_{i-1}}^{t_i}; X_0^{t_i} \big| Y_0^{t_{i-1}}, V).
    \label{eq: mut inf summands in definition of ieps}
\end{align}

We define
$\Tc(a,b)$ to be the set of all finite partitions of the time
interval $[a,b)$. The quantity
$I_{\tv} ( X_0^T \to Y_0^T )$ is monotone  in $\tv$ in the following
sense:
\begin{proposition} \label{claim: monotonicity of It}
Let  $\tv$ and  $\mathbf{t'}$ be partitions in
 $\Tc(0,T)$.
If $\mathbf{t'}$ is a refinement of $\tv$, i.e., $\{t_i\} \subset \{t'_i\}$,
then
$I_{\mathbf{t'}} ( X_0^T \to Y_0^T ) \le
I_{\tv} ( X_0^T \to Y_0^T )$.
\end{proposition}

\begin{IEEEproof} It suffices to prove the claim assuming $\tv$
as in \eqref{eq: what components of t satisfy} and that  $\mathbf{t'}$
is the $(n+2)$-dimensional vector with components
\begin{equation}\label{eq: what components of t' satisfy}
    0 = t_0 < t_1 < \cdots < t_{i-1} < t' < t_i < \cdots < t_n = T.
\end{equation}
For such $\tv$ and  $\mathbf{t'}$, we have
from \eqref{eq: mut inf summands in definition of ieps}
\begin{align}
&I_\tv ( X_0^T \to Y_0^T ) - I_{\mathbf{t'}} ( X_0^T \to Y_0^T ) \\
&\qquad =   I ( Y_{t_{i-1}}^{t_i} ; X_0^{t_i} | Y_0^{t_{i-1}}
   )  - \bigl[ I ( Y_{t_{i-1}}^{t' } ; X_0^{t' } | Y_0^{t_{i-1}}
   ) + I ( Y_{t'}^{t_i } ; X_0^{t_i } | Y_0^{t'}
   ) \bigr] \\
&\qquad = I ( Y_{t_{i-1}}^{t_i } ; X_0^{t_i } | Y_0^{t_{i-1}}
   )  - \bigl[ I ( Y_{t_{i-1}}^{t' } ; X_0^{t' } | Y_0^{t_{i-1}}
   ) + I ( Y_{t'}^{t_i } ; X_0^{t_i } |
   Y_0^{t_{i-1}} , Y_{t_{i-1}}^{t' }
   ) \bigr] \\
&\qquad=  I ( X_0^{t'}, X_{t'}^{t_i}; Y_{t_{i-1}}^{t'}, Y_{t'}^{t_i}
                   | Y_0^{t_{i-1}})
        - \bigl[ I ( Y_{t_{i-1}}^{t' }; X_0^{t' } | Y_0^{t_{i-1}} )
        + I ( Y_{t'}^{t_i}; X_0^{t'}, X_{t'}^{t_i }
                   | Y_0^{t_{i-1}} , Y_{t_{i-1}}^{t'} ) \bigr] \\
&\qquad = I ( X_0^{t'}, X_{t'}^{t_i }; Y_{t_{i-1}}^{t'}, Y_{t'}^{t_i } | Y_0^{t_{i-1}}
   ) - I ( X_0^{t'}  X_{t'}^{t_i } \to Y_{t_{i-1}}^{t'}, Y_{t'}^{t_i } | Y_0^{t_{i-1}}
   ) \\
&\qquad \ge 0,
\end{align}
where the last inequality follows since directed
information (between two sequences of length 2 in this case) is upper bounded by the mutual information \cite[Th.~2]{Massey90}.
\end{IEEEproof}

The following definition is now natural:
\begin{definition} \label{def: definition of directed info in cont
time} Let $(X_0^{T} , Y_0^{T} )$ be a pair of stochastic processes.
The
\emph{directed information} from $X_0^T$ to $Y_0^T$ is
defined as
\begin{equation}\label{eq: directed info in continuous time defined}
I (  X_0^T \to Y_0^T ) := \inf_{\tv \in \Tc(0,T)}
I_{\tv} (  X_0^T \to Y_0^T ).
\end{equation}
If $V$ is another random object jointly distributed with $(X_0^T, Y_0^T)$ we define the conditional directed information  $I (  X_0^T \to Y_0^T | V )$ as
\begin{equation}\label{eq: directed info in continuous time defined cond}
I (  X_0^T \to Y_0^T | V ) := \inf_{\tv \in \Tc(0,T)}
I_{\tv} (  X_0^T \to Y_0^T | V ).
\end{equation}
\end{definition}
Note that the definitions and conventions preceding Definition \ref{def: definition of directed info in cont
time} imply that the directed information $I (  X_0^T \to Y_0^T )$ is 
a nonnegative extended real number (i.e., as an element of $[0, \infty]$).
It is also worth noting, by recalling  \eq{eq: mutinf defined ALT}, that each of the conditional mutual information terms in \eq{eq: mut inf summands in definition of ieps}, and hence the sum, is a supremum over ``space'' partitions of the
stochastic process in the corresponding time intervals. Thus the directed information in \eq{eq: directed info in continuous time defined} is an infimum over time partitions of a supremum over space partitions.

Also note that
\begin{equation}\label{eq: alternative def or relation}
    I (  X_0^T \to Y_0^T )
    = \lim_{\eps \to  0^+} \inf_{\tv: t_i - t_{i-1}  \le \eps, \forall i}
I_{\tv} (  X_0^T \to Y_0^T ),
\end{equation}
where the infimum is over all partitions in $\Tc(0,T)$ with subinterval lengths uniformly bounded by $\epsilon>0$.
Indeed, for any $\epsilon>0$ and any partition $\tv \in \Tc(0,T)$, have
$\inf_{\tv': t_i' - t_{i-1}'  \le \eps, \forall i}I_{\tv'} (  X_0^T \to Y_0^T ) \leq
I_{\tv} (  X_0^T \to Y_0^T )$,
since a  refinement of  the time interval does not increase the directed information
as seen in Proposition \ref{claim: monotonicity of It}.
By the arbitrariness of $\tv \in \Tc(0,T)$, this implies
\begin{equation}
\inf_{\tv': t_i' - t_{i-1}'  \le \eps, \forall i}I_{\tv'} (  X_0^T \to Y_0^T ) \leq
 \inf_{\tv \in \Tc(0,T)} I_{\tv} (  X_0^T \to Y_0^T ) = I (  X_0^T \to Y_0^T ),
\end{equation}
which in turn implies $ I (  X_0^T \to Y_0^T )
    \geq \lim_{\eps \to  0^+} \inf_{\tv: t_i - t_{i-1}  \le \eps, \forall i}
I_{\tv} (  X_0^T \to Y_0^T )$ by the arbitrariness of $\eps > 0$. Since the reverse inequality  $ I (  X_0^T \to Y_0^T )
    \leq \lim_{\eps \to  0^+} \inf_{\tv: t_i - t_{i-1}  \le \eps, \forall i}
I_{\tv} (  X_0^T \to Y_0^T )$ is immediate from the definition of  $I (  X_0^T \to Y_0^T )$, we have \eqref{eq: alternative def or relation}.


As is clear from its definition in \eqref{eq: directed info in discrete time defined},     the discrete-time directed information satisfies
\begin{equation}
I(X^n \to Y^n) - I(X^{n-1} \to Y^{n-1}) =  I( Y_n ; X^n |Y^{n-1}).
\end{equation}
A continuous-time analogue would be that, for small $\delta > 0$,
\begin{equation} \label{eq: approximate relation in cont time}
I (  X_0^{t+ \delta} \to Y_0^{t+ \delta} ) - I (  X_0^t \to Y_0^t ) \approx   I(Y_t^{t+\delta};
X_0^{t+\delta}|Y_0^t).
\end{equation}
Thus,  if our proposed notion of directed information in continuous time is to be a natural extension of that in discrete time, one might expect
the approximate relation \eqref{eq: approximate relation in cont time} to hold in some sense. Toward a precise statement, denote
\begin{equation}
i_t :=   \lim_{\delta
\to 0^+} \frac{1}{\delta} I(Y_t^{t+\delta}  ;
X_0^{t+\delta}|Y_0^t) \quad \text{for } t \in (0,T)
\label{eq: it defined}
\end{equation}
whenever the limit exists.  Assuming $i_t$ exists, let
\begin{equation}
\eta (t , \delta) :=  \frac{1}{\delta} I(Y_t^{t+\delta}  ;
X_0^{t+\delta}|Y_0^t) - i_t
\label{eq: it defined and equivaleted defined}
\end{equation}
and note that \eqref{eq: it defined} is equivalent  to
\begin{equation}
\lim_{\delta
\to 0^+}  \eta (t, \delta) =  0 .
\label{eq: it defined and equivaleted}
\end{equation}

\begin{proposition}
 \label{claim: dir inf in terms of its differential}
Fix $0 < t < T$. Suppose that $i_t$ is continuous at $t$ and
that the convergence in \eqref{eq: it defined and equivaleted}
is uniform in the interval $[t, t+\gamma)$ for some $\gamma > 0$.
Then
\begin{equation}\label{eq: the differential for every t}
    \frac{d^{+}}{dt} I (  X_0^t \to Y_0^t ) =  i_t .
\end{equation}
\end{proposition}
Note that Proposition \ref{claim: dir inf in terms of its differential} formalizes
\eqref{eq: approximate relation in cont time} by implying that the left and right hand sides of  \eqref{eq: approximate relation in cont time}, when normalized by $\delta$, coincide in the limit of small $\delta$.

\begin{IEEEproof}[Proof of Proposition \ref{claim: dir inf in terms of its differential}]
Note first that the stipulated uniform convergence
in \eqref{eq: it defined and equivaleted} implies the existence of $\gamma > 0$ and  a monotone function  $f(\delta)$  such that
\begin{equation}\label{eq: f as bound on eta}
    | \eta (t' , \delta) | \le f(\delta)  \quad\text{for all } t' \in [t, t+ \gamma )
\end{equation}
and
\begin{equation}\label{eq: f is vanishing}
   \lim_{\delta
\to 0^+}  f(\delta) = 0.
\end{equation}
Fix now $0 < \eps \le \gamma$ and consider
\begin{align}
I ( X_0^{t + \eps} \to Y_0^{t + \eps} )
&= \inf_{\tv \in \Tc(0,t+ \eps)}
I_{\tv} (  X_0^{t + \eps} \to Y_0^{t + \eps} ) \\
&= \inf_{\tv \in \Tc(0,t+ \eps)}  \sum_{i =1}^n I ( Y_{t_{i-1}}^{t_i} ; X_0^{t_i} | Y_0^{t_{i-1}}   )  \\
&\stackrel{}{=} \inf_{\tv \in \left(\Tc(0,t)\bigcup \Tc(t,t+ \eps)\right)}  \sum_{i =1}^n I ( Y_{t_{i-1}}^{t_i} ; X_0^{t_i} | Y_0^{t_{i-1}}   )  \label{e_interval_t}\\
&= \inf_{\tv \in \Tc(0,t)}   \sum_{i = 1}^n I ( Y_{t_{i-1}}^{t_i} ; X_0^{t_i} | Y_0^{t_{i-1}} )
  +  \inf_{\tv \in \Tc(t,t+ \eps)}
   \sum_{i = 1}^n I ( Y_{t_{i-1}}^{t_i}; X_0^{t_i} | Y_0^{t_{i-1}} )  \\
&=  I (  X_0^{t } \to Y_0^{t } ) +   \inf_{\tv \in \Tc(t,t+ \eps)}  \sum_{i = 1}^n ( t_i - t_{i-1}) \frac{1}{t_i - t_{i-1}} I ( Y_{t_{i-1}}^{t_i} ; X_0^{t_i} | Y_0^{t_{i-1}} ) \\
&=  I (  X_0^{t } \to Y_0^{t } )
  + \inf_{\tv \in \Tc(t,t+ \eps)}
    \sum_{i = 1}^n ( t_i - t_{i-1}) \cdot [ i_{t_{i-1}}
    + \eta (  t_{i-1} , t_i - t_{i-1})   ],
\label{eq: directed inf plus eps}
\end{align}
where the equality
in \eqref{e_interval_t}  follows since the infimum over all partitions does not change by restricting to partitions that have an interval up to time $t$ and from time $t$
and the last equality follows by the definition of
the function $\eta$ in \eqref{eq: it defined and equivaleted defined}.
Now,
\begin{align}
\inf_{\tv \in \Tc(t,t+ \eps)}  \sum_{i = 1}^n ( t_i - t_{i-1}) \cdot
[ i_{t_{i-1}} + \eta (  t_{i-1} , t_i - t_{i-1})   ]
& \le
   \inf_{\tv \in \Tc(t,t+ \eps)}  \sum_{i = 1}^n ( t_i - t_{i-1}) \cdot
   \biggl[\,\sup_{t' \in [t , t+ \eps)} i_{t'} + f (\eps)   \biggr]
   \label{eq: exp of int with eta terms} \\
& =   \eps \biggl[\, \sup_{t' \in [t , t+ \eps)} i_{t'} + f (\eps) \biggr],
   \label{eq: upper bound on the diff}
\end{align}
where the inequality in \eqref{eq: exp of int with eta terms}
is due to \eqref{eq: f as bound on eta} and the monotonicity of $f$,
 which implies $f ( t_i - t_{i-1}) \le f (\eps )$, as $t_i - t_{i-1}$ is the length of a subinterval in $[t, t + \eps)$.  Bounding the $\eta$ terms in \eqref{eq: exp of int with eta terms} from the other direction,  we similarly obtain
\begin{equation}
   \inf_{\tv \in \Tc(t,t+ \eps)}  \sum_{i = 1}^n ( t_i - t_{i-1}) \cdot [ i_{t_{i-1}} + \eta (  t_{i-1} , t_i - t_{i-1})   ] \ge
      \eps \biggl[\, \inf_{t' \in [t , t+ \eps)} i_{t'} - f (\eps)   \biggr].
       \label{eq: lower bound on the diff}
\end{equation}
Combining \eqref{eq: directed inf plus eps},
\eqref{eq: upper bound on the diff}, and \eqref{eq: lower bound on the diff}
yields
\begin{equation}
\inf_{t' \in [t , t+ \eps)} i_{t'} - f (\eps)  \le \frac{ I (  X_0^{t + \eps} \to Y_0^{t + \eps} )  - I (  X_0^{t } \to Y_0^{t } ) }{\eps}  \le
\sup_{t' \in [t , t+ \eps)} i_{t'} + f (\eps)  \quad \text{for all } \eps > 0.
\label{eq: non eps limit u and a}
\end{equation}
The continuity of $i_t$ at $t$ implies
$\lim_{\eps \to 0^+} \inf_{t' \in [t , t+ \eps)} i_{t'}  = \lim_{\eps \to 0^+} \sup_{t' \in [t , t+ \eps)} i_{t'} = i_t$ and thus,
 taking the limit $\eps \to 0^+$ in \eqref{eq: non eps limit u and a} and
applying \eqref{eq: f is vanishing} finally yields
\begin{equation}
\lim_{\eps \to 0^+}  \frac{ I (  X_0^{t + \eps} \to Y_0^{t + \eps} )  - I (  X_0^{t } \to Y_0^{t } ) }{\eps} = i_t,
\end{equation}
which completes the proof of Proposition \ref{claim: dir inf in terms of its differential}.
\end{IEEEproof}

Beyond the intuitive appeal of Proposition \ref{claim: dir inf in terms of its differential} in formalizing  \eqref{eq: approximate relation in cont time}, it also provides a useful formula for computing directed information. Indeed, the integral version of \eqref{eq: the differential for every t} is
\begin{equation}\label{eq: the differential for every t integral vers}
     I (  X_0^T \to Y_0^T ) =  \int_0^T i_t  \, dt.
\end{equation}
As the following example illustrates, evaluating the right hand side of \eqref{eq: the differential for every t integral vers} (via the definition of $i_t$ in \eqref{eq: it defined}) can be simpler  than tackling the left hand side directly via Definition \ref{def: definition of directed info in cont
time}.

\begin{example}
Let $\{B_t\}$ be a standard Brownian motion and $A \sim
\mathrm{N}(0,1)$ be independent of $\{B_t\}$. Let $X_t \equiv A$ for
all $t$ and $dY_t = X_t dt + dB_t$. Letting $J(P, N)
= (1/2) \ln ((P+N)/N)$ denote
the mutual information between a Gaussian random variable of variance
$P$ and its corrupted version by an independent Gaussian noise of
variance $N$, we have for every $t \in [0,T)$
\begin{equation*}\label{eq: the differential in gaussian case}
I(Y_t^{t+\delta}; X_0^{t+\delta}|Y_0^t)
= J \left( \frac{1/t}{1+1/t}, \frac{1}{\delta} \right)
= \frac{1}{2} \ln \left( 1 + \frac{\delta}{t+1} \right) .
\end{equation*}
With such an explicit expression for $I(Y_t^{t+\delta}; X_0^{t+\delta}|Y_0^t)$,  $i_t$ can be obtained directly from its definition:
\begin{equation}
i_t
=
 \lim_{\delta \to 0^+} \frac{1}{2 \delta}
\ln \left( 1 + \frac{\delta}{t+1} \right)
= \frac{1}{2 (t+1)}.
\end{equation}
We can now compute the directed information by applying
Proposition~\ref{claim: dir inf in terms of its differential}:
\begin{equation}
I (  X_0^T \to Y_0^T ) = \int_0^T i_t  dt
= \int_0^T \frac{1}{2 (t+1)} dt =
\frac{1}{2} \ln (1 + T). \label{eq:applying-claim-for-diff-int-for-example}
\end{equation}
Note that in this example $I ( X_0^T ; Y_0^T ) = J (1, 1/T)
= \frac{1}{2} \ln (1 + T)$ and thus, by
\eqref{eq:applying-claim-for-diff-int-for-example}, we have $I (
X_0^T \to Y_0^T ) = I ( X_0^T ; Y_0^T
)$. This equality between mutual information and directed information holds
in more general situations, as elaborated in the next section.
\end{example}

The directed information we have just defined is between two
processes on $[0,T)$. We extend this definition  to processes of different durations
by zero-padding at the beginning of the shorter process. For instance,
\begin{equation} \label{eq: T-delta directed info}
I ( X_0^{T-\delta} \to Y_0^T ) := I(  (0_0^\delta X_0^{T-\delta})
\to Y_0^T ),
\end{equation}
where $(0_0^\delta X_0^{T-\delta})$ denotes a process on $[0,T)$ formed
by concatenating a
process that is equal to the constant $0$ for the time interval $[0,\delta)$
and then the process $X_0^{T-\delta}$.

Define now
\begin{equation}\label{eq: limsup dir inf for t-}
 \overline{I} (  X_0^{T-} \to Y_0^T ) := \limsup_{\delta \to 0^+} I (  X_0^{T-\delta} \to Y_0^T )
\end{equation}
and
\begin{equation}\label{eq: liminf dir inf for t-}
 \underline{I} (  X_0^{T-} \to Y_0^T ) := \liminf_{\delta \to 0^+} I (  X_0^{T-\delta} \to Y_0^T
 ).
\end{equation}
 Finally, define the
directed information $I ( X_0^{T-} \to Y_0^T )$
by
\begin{equation}\label{eq: the notion of dir info for a slightly left shifted process}
    I ( X_0^{T-} \to Y_0^T ) := \lim_{\delta \to 0^+} I (
X_0^{T-\delta} \to Y_0^T
 )
\end{equation}
when the limit exists, or equivalently,
when $\overline{I} ( X_0^{T-} \to Y_0^T ) = \underline{I} ( X_0^{T-} \to Y_0^T )$. As we shall see below (in the last part of Proposition \ref{prop: 2 properties of directed info}),  $I ( X_0^{T-} \to Y_0^T )$ is guaranteed to exist whenever $I(X_0^T ; Y_0^T) < \infty$.

\section{Properties of the Directed Information in Continuous Time}
\label{sec: Properties of the Directed Information in Continuous
Time} The following proposition collects some properties of directed
information in continuous time:
\begin{proposition} \label{prop: 2 properties of directed info}
Let $(X_0^T , Y_0^T )$ be a pair of jointly distributed stochastic processes.
Then:
\begin{enumerate}
  \item Monotonicity: $I(X_0^t \to Y_0^t)$ is monotone
  nondecreasing in $0 \leq t \leq T$.

  \item Invariance to time dilation: For $\alpha > 0$, if $\tilde{X}_t = X_{t
  \alpha}$ and $\tilde{Y}_t = Y_{t
  \alpha}$, then
  $I ( \tilde{X}_0^{T/\alpha } \to \tilde{Y}_0^{T/\alpha}
  ) = I(X_0^T \to Y_0^T)$.
  More generally, if $\phi$ is
  monotone strictly increasing and continuous, and $(\tilde{X}_{\phi(t)} , \tilde{Y}_{\phi(t)}) = (X_t ,
  Y_t)$, then
\begin{equation}\label{eq: general invariance principle}
I(X_0^T \to Y_0^T) = I ( \tilde{X}_{\phi(0)}^{\phi(T)}
\to \tilde{Y}_{\phi(0)}^{\phi(T)}
  ) .
\end{equation}

  \item Coincidence of directed information and mutual information: If the Markov relation
  $Y_0^t \to X_0^t \to X_t^T$ holds for all $0 \le  t < T$,
  then
\begin{equation}\label{eq: dir and mut inf equal when markov}
    I (  X_0^T \to Y_0^T ) = I (  X_0^T ; Y_0^T ).
\end{equation}

\item Equivalence between discrete time and piecewise constancy in
continuous time: Let $(U^n, V^n)$ be a pair of jointly distributed
$n$-tuples and suppose $(t_0, t_1, \ldots, t_n)$ satisfy~\eqref{eq: what components of t satisfy}.
Let the pair $(X_0^T, Y_0^T)$ be defined as the
piecewise-constant process satisfying
\begin{equation}\label{eq: construction of the piecewise constant process}
    (X_t,Y_t) = (U_i, V_i)   \quad\text{if } t_{i-1} \le t < t_i
\end{equation}
for $i = 1,\ldots, n$. Then
\begin{equation}
I (  X_0^T \to Y_0^T )
= I (  U^n \to V^n ).
\end{equation}

\item Conservation law: For any $0 < \delta \le T$ we have
\begin{equation} \label{eq: conservation law in cont time}
I(X_0^\delta; Y_0^\delta) + I( X_0^T \to Y_\delta^T | Y_0^\delta) +
I ( Y_0^{T-\delta} \to X_0^T ) = I (  X_0^T ; Y_0^T ).
\end{equation}
Further, if $ I (  X_0^T ; Y_0^T ) < \infty$ then
 $I ( Y_0^{T-} \to
X_0^T )$ exists and
\begin{equation} \label{eq: conservation law under continuity}
I ( X_0^T \to Y_0^T )  + I ( Y_0^{T-} \to X_0^T )
= I (  X_0^T ; Y_0^T ).
\end{equation}
\end{enumerate}
\end{proposition}

\begin{remarks}
\mbox{}
\begin{enumerate}
  \item  The first, second, and fourth parts in the proposition present properties
  that are known to hold for mutual information (when all the directed information expressions in those items are replaced by the corresponding
  mutual information), which follow immediately from the data
  processing inequality and the invariance of mutual
  information to one-to-one transformations of its arguments.    That these properties hold also for
   directed information is not as obvious in view of the fact that directed
information is, in general, not invariant to one-to-one
transformations nor does it satisfy the data processing inequality in
its second argument.

\item The third part of the proposition is a natural analogue of
  the fact that $I(X^n ; Y^n) = I(X^n \to Y^n)$ whenever
  $Y^i \to X^i \to X_{i+1}^n$ form a Markov chain for all $1 \le i \le n$. It
covers, in particular, any scenario where $X_0^T$ and $Y_0^T$ are the input and output of any
channel of the form $Y_t = g_t (X_0^t, W_0^T)$, where the process $W_0^T$ (which can be thought
of as the internal channel noise) is independent of the channel input process $X_0^T$. To see
this,
note that
in this case we have $(X_0^t, W_0^T) \to X_0^t \to X_t^T$ for all $0
\le  t \le T$, implying $Y_0^t \to X_0^t \to X_t^T$
since $Y_0^t$ is determined by the pair $(X_0^t, W_0^T)$.

  \item Particularizing even further, we obtain
  $I (  X_0^T \to Y_0^T ) = I (  X_0^T ; Y_0^T )$
  whenever $Y_0^T$ is the outcome of corrupting $X_0^T$ with  additive
  noise, i.e., $Y_t = X_t + W_t$, where $X_0^T$ and $W_0^T$ are
  independent.


\item The fifth part of the proposition can be considered the
continuous-time analogue of the discrete-time conservation law \cite{Massey05}
\begin{equation}\label{eq: conservation law for discrete case}
    I(U^n \to V^n)  + I(V^{n-1} \to U^n)  = I(U^n ; V^n).
\end{equation}
It is consistent with, and in fact generalizes, the third part. Indeed, if the Markov relation
$Y_0^t \to X_0^t \to X_t^T$ holds
for all $0 \le  t \le T$ then our definition of  directed
information is readily seen to imply that $I ( Y_0^{T-\delta} \to X_0^T ) = 0$
for all $\delta > 0$ and therefore that $I ( Y_0^{T-} \to X_0^T )$ exists and
equals zero. Thus \eqref{eq: conservation law under continuity} in this case reduces to
\eqref{eq: dir and mut inf equal when markov}.
\end{enumerate}
\end{remarks}

\begin{IEEEproof}[Proof  of Proposition \ref{prop: 2 properties of directed info}]
The first part of the proposition follows immediately
from the definition of directed information in continuous time
(Definition~\ref{def: definition of directed info in cont
time})
and from the
fact that, in discrete time, $I(U^m \to V^m) \le  I(U^n
\to V^n)$ for $m \le n$. The second part follows from
Definition~\ref{def: definition of directed info in cont
time} upon noting that, under a dilation
$\phi$ as stipulated, due to the invariance of mutual information to
one-to-one transformations of its arguments, for any
partition $\tv$ of $[0,T)$,
\begin{equation} \label{eq: It and Iphioft}
I_{\tv} (  X_0^T \to Y_0^T ) =
I_{\phi(\tv)} ( \tilde{X}_{\phi(0)}^{\phi(T)}
\to \tilde{Y}_{\phi(0)}^{\phi(T)} ),
\end{equation}
where $\phi(\tv)$ is shorthand for $(\phi(t_0, \phi(t_1),
\ldots, \phi(t_n) )$. Thus
\begin{align}
I (  X_0^T \to Y_0^T )
&= \inf_{\tv \in \Tc(0,T)}
   I_{\tv} (  X_0^T \to Y_0^T )
   \label{eq: property 2 eq a}\\
&= \inf_{\tv \in \Tc(0,T)}
   I_{\phi(\tv)} ( \tilde{X}_{\phi(0)}^{\phi(T)}
                               \to \tilde{Y}_{\phi(0)}^{\phi(T)} )
  \label{eq: property 2 eq b}\\
&= \inf_{\tv \in \Tc(\phi(0),\phi(T))}
   I_{\tv} ( \tilde{X}_{\phi(0)}^{\phi(T)}
                         \to \tilde{Y}_{\phi(0)}^{\phi(T)} )
  \label{eq: property 2 eq c}\\
& = I ( \tilde{X}_{\phi(0)}^{\phi(T)}
            \to \tilde{Y}_{\phi(0)}^{\phi(T)}   ),
  \label{eq: property 2 eq d}
\end{align}
where \eqref{eq: property 2 eq a} and~\eqref{eq: property 2 eq d}
follow from Definition \ref{def: definition of directed info in cont time},
\eqref{eq: property 2 eq b} follows from \eqref{eq: It and Iphioft},
and \eqref{eq: property 2 eq c} is due to the
strict monotonicity and continuity of $\phi$ which implies that
\begin{equation}
\{ \phi(\tv) : \tv \mbox{ is a partition of }  [0,T) \}
= \{ \tv : \tv \mbox{ is a partition of }  [\phi(0),\phi(T)) \}.
\end{equation}

Moving to the proof of the third part, assume that the Markov relation
$Y_0^t \to X_0^t \to X_t^T$ holds for all $0 \le  t \le T$
and fix $\tv = (t_0, t_1, \ldots, t_n )$ as in
\eqref{eq: what components of t satisfy}. Then
\begin{align}
I_{\tv} (  X_0^T
\to Y_0^T )
&= I( X_0^{T, \tv} \to Y_0^{T, \tv} ) \label{eq: directed info between sequences II} \\
&= \sum_{i=1}^N I( Y_{t_{i-1}}^{t_i }; X_0^{t_i } | Y_0^{t_{i-1}} ) \\
&= \sum_{i=1}^N I( Y_{t_{i-1}}^{t_i }; X_0^T | Y_0^{t_{i-1}} )
   \label{eq: in pf follows from markov struct} \\
&= I(  X_0^T ; Y_0^T ),
    \label{eq: chain rule  for entropy}
\end{align}
where \eqref{eq: in pf follows from markov struct} follows
since $Y_0^{t_i} \to X_0^{t_i} \to
X_{t_i}^T$ for each $1 \le  i \le N$, and \eqref{eq: chain rule
for entropy} is due to the chain rule for mutual information. The proof of the third
part of the proposition now follows from the arbitrariness of $\tv$.

To prove the fourth part, consider first the case $n=1$. In this
case $X_t \equiv U_1$ and $Y_t \equiv V_1$ for all $t \in [0,T)$. It
is an immediate consequence of the definition of directed
information that $I((U, U, \ldots, U) \to (V,V, \ldots, V))
= I(U;V)$ and therefore that $I_{\tv} (  X_0^T
\to Y_0^T ) = I (U_1 ; V_1) = I(U_1 \to V_1)$
for all $\tv$. Consequently $I (  X_0^T \to
Y_0^T ) = I(U_1 \to V_1)$, which establishes the case
$n=1$. For the general case $n \ge 1$, note first that it
is immediate from the definition of $I_{\tv} (  X_0^T
\to Y_0^T )$ and from the construction of  $(X_0^T,
Y_0^T)$ based on $( X^n , Y^n)$   in \eqref{eq: construction of the
piecewise constant process} that for $\tv = (t_0, t_1,
\ldots, t_n )$ consisting of the time epochs in \eqref{eq: construction
of the piecewise constant process} we have  $I_{\tv} (
X_0^T \to Y_0^T ) =  I ( U^n \to V^n
  )$. Thus $I (  X_0^T \to Y_0^T ) \le I_{\tv} (  X_0^T
\to Y_0^T ) =  I ( U^n \to V^n
  )$. We now argue that
\begin{equation}\label{eq: lower bounding the dir inf induced by a partition}
I_{\mathbf{s}} (  X_0^T \to Y_0^T ) \ge I ( U^n \to V^n )
\end{equation}
for any partition $\mathbf{s}$. By Proposition~\ref{claim: monotonicity of It},
it suffices to establish \eqref{eq: lower bounding the dir inf induced by a partition} with equality assuming $\mathbf{s}$ is a refinement
of  the particular $\tv$ just discussed, that is,
$\mathbf{s}$ is of the form
\begin{equation}\label{eq: what components of s satisfy}
0 = t_0 = s_{0,0} < s_{0,1} < \cdots < s_{0,J_0} < t_1 = s_{1,0}
< s_{1,1} < \cdots < s_{1, J_1} < t_2 = s_{2,0}
< \cdots  < s_{n-1, J_{n-1}} < t_n = T.
\end{equation}
Then,
\begin{align}
I_{\mathbf{s}} (  X_0^T \to Y_0^T )
&= I ( X_0^{T, \mathbf{s}} \to Y_0^{T, \mathbf{s}} ) \label{eq: directed info between sequences ?} \\
&=   \sum_{i=1}^n \sum_{j=1}^{J_{i-1}} I ( Y_{s_{i-1, j-1}}^{s_{i-1,j} } ; X_0^{s_{i-1,j} } | Y_0^{s_{i-1,j-1}}    )
   \label{eq: 2nd hald of monst in 2part prop}\\
&= \sum_{i=1}^n I(U_i ; V^i | U^{i-1})
   \label{eq: in pf follows from markov struct ?} \\
&= I (U^n \to V^n ), \label{eq: chain rule  for entropy ?}
\end{align}
where \eqref{eq: in pf follows from markov struct ?} follows by applying a similar argument as in the case $n=1$.

Moving  to the proof of the fifth part of the proposition, fix
$\tv = (t_0, t_1, \ldots, t_n )$ as in \eqref{eq: what
components of t satisfy} with $t_1 =\delta > 0$. Applying the
discrete-time conservation law \eqref{eq: conservation law for discrete
case}, we have
\begin{equation} \label{eq: discrete time cons law on chopped proc}
I_{\tv} ( X_0^T \to Y_0^T )
+ I_{\tv} ( Y_0^{T-\delta} \to X_0^T )
= I ( X_0^T ; Y_0^T )
\end{equation}
and consequently, for any $\eps > 0$,
\begin{align}
& \inf_{ \{ \tv: t_1 = \delta,  \max_{i \ge 2} t_i - t_{i-1}
\le \eps \}} I_{\tv} (  X_0^T \to Y_0^T ) + \inf_{ \{ \tv:
\max_{i} t_i - t_{i-1} \le \eps \}} I_{\tv} ( Y_0^{T-\delta} \to X_0^T
)
\label{eq: taking now arbit t with delta} \\
&\qquad= \inf_{ \{ \tv: t_1 = \delta,  \max_{i \ge 2} t_i - t_{i-1}
\le \eps \}} I_{\tv} (  X_0^T \to Y_0^T ) + \inf_{ \{ \tv: t_1
= \delta,  \max_{i \ge 2} t_i - t_{i-1} \le \eps \}} I_{\tv} ( Y_0^{T-\delta}
\to X_0^T )
\label{eq: taking now arbit t with delta step a} \\
&\qquad= \inf_{ \{ \tv: t_1 = \delta,  \max_{i \ge 2} t_i -
t_{i-1} \le \eps \}} \bigl[ I_{\tv} (  X_0^T \to Y_0^T ) +
I_{\tv} ( Y_0^{T-\delta}
\to X_0^T ) \bigr] \label{eq: in dirminus by refinement prop} \\
&\qquad= I ( X_0^T ; Y_0^T ),
\label{eq: in dirminus by refinement prop eps pos}
\end{align}
where
the equality in \eqref{eq: taking now arbit t with delta step a}
follows since due to its
definition in \eqref{eq: T-delta directed info},
$I_{\tv} ( Y_0^{T-\delta} \to X_0^T )$ does not decrease by
refining the time interval $\tv$ in the $[0,
  \delta)$ interval;
the equality in \eqref{eq: in dirminus by refinement prop} follows from the
refinement property in Proposition~\ref{claim: monotonicity of It}, which
implies that for arbitrary processes $X_0^T, Y_0^T, Z_0^T, W_0^T $
and partitions $\tv$ and $\tv'$ there exists a third
partition $\tv''$ (which will be a refinement of both) such
that
\begin{equation}\label{eq: following form refineprop}
    I_{\tv} (  X_0^T \to Y_0^T ) + I_{\tv'} (  Z_0^T \to W_0^T
    ) \ge I_{\tv''} (  X_0^T \to Y_0^T ) + I_{\tv''} (  Z_0^T \to W_0^T );
\end{equation}
and the equality in \eqref{eq: in dirminus by refinement prop eps pos}
follows since
\eqref{eq: discrete time cons law on chopped proc} holds
for any
$\tv = (t_0, t_1, \ldots, t_n)$
with $t_1 =\delta$.
Hence,
\begin{align}
I ( X_0^T ; Y_0^T )
&=  \lim_{\eps \to 0^+} \biggl[\, \inf_{ \{ \tv: t_1 = \delta,  \max_{i \ge 2} t_i - t_{i-1} \le
\eps \}} I_{\tv} (  X_0^T \to Y_0^T ) + \inf_{ \{ \tv:  \max_{i
} t_i - t_{i-1} \le \eps \}} I_{\tv} ( Y_0^{T-\delta} \to
X_0^T ) \biggr]
\label{eq: property 5 step a}  \\
&= \lim_{\eps \to 0^+}  \inf_{ \{ \tv: t_1 = \delta,  \max_{i \ge 2} t_i - t_{i-1}
\le \eps \}} I_{\tv} (  X_0^T \to Y_0^T ) + \lim_{\eps \to
0^+} \inf_{ \{ \tv:  \max_{i } t_i - t_{i-1} \le \eps \}} I_{\tv} (
Y_0^{T-\delta} \to
X_0^T )  \label{eq: frist exp w 2 limits} \\
&=  \lim_{\eps \to 0^+}  \inf_{ \{ \tv: t_1 = \delta,  \max_{i \ge 2} t_i - t_{i-1}
\le \eps \}} \Biggl[  I(X_0^{\delta} ; Y_0^{\delta}) +  \sum_{i=2}^{n} I (
Y_{t_{i-1}}^{t_i } ; X_0^{t_i } | Y_0^{t_{i-1}}
   ) \Biggr] +  I ( Y_0^{T-\delta} \to
X_0^T ) \label{eq: property 5 step b} \\
&= I(X_0^{\delta} ; Y_0^{\delta}) + \lim_{\eps \to 0^+}  \inf_{ \{ \tv: t_1 = \delta,  \max_{i \ge 2} t_i - t_{i-1}
\le \eps \}}      \sum_{i=2}^{n} I ( Y_{t_{i-1}}^{t_i } ; X_0^{t_i } |
Y_0^{t_{i-1}})
+  I ( Y_0^{T-\delta} \to X_0^T )  \\
&= I(X_0^{\delta} ; Y_0^{\delta}) + I( X_0^T \to  Y_\delta^T | Y_0^{\delta})
+  I ( Y_0^{T-\delta} \to X_0^T ),
\label{eq: property 5 step c}
\end{align}
where
the equality in~\eqref{eq: property 5 step a}
 follows by taking the limit $\eps \to 0$ from both sides of
\eqref{eq: in dirminus by refinement prop eps pos};
the equality in~\eqref{eq: property 5 step b} follows
by writing out $I_{\tv} (  X_0^T \to Y_0^T
  )$ explicitly for $\tv$ with $t_1 = \delta$ and
  using  \eqref{eq: alternative def or relation} to equate the second
  limit in \eqref{eq: frist exp w 2 limits} with $I ( Y_0^{T-\delta} \to
X_0^T )$;
and the equality in~\eqref{eq: property 5 step c}
follows by applying \eqref{eq: alternative def or
  relation} on the conditional distribution of the pair $(X_0^T , (0_0^\delta Y_\delta^T))$ given $Y_0^\delta$.
We have thus proven \eqref{eq: conservation law in cont time} or,
equivalently, the identity
\begin{equation}\label{eq: identity equiv to cons law delta}
I(X_0^{\delta} ; Y_0^{\delta}) + I( X_0^T \to Y_\delta^T | Y_0^{\delta}) = I
(  X_0^T ; Y_0^T
  ) - I ( Y_0^{T-\delta}
\to X_0^T ) .
\end{equation}
Toward the proof of \eq{eq: conservation law under continuity}, for  $\tv \in \Tc(0,T)$ and $\delta < t_1$
 let  $\tv_\delta$ denote the refinement of  $\tv$ obtained by adding an additional point at $\delta$.  Then
 \begin{eqnarray}
  I_{\tv} (  X_0^T
\to Y_0^T ) & \geq & I_{\tv_\delta} (  X_0^T
\to Y_0^T ) \\ &  = & I(X_0^{\delta} ; Y_0^{\delta}) + I_{\tv_\delta}( X_0^T \to Y_\delta^T | Y_0^{\delta}) \\ &  \geq &  I(X_0^{\delta} ; Y_0^{\delta}) + I( X_0^T \to Y_\delta^T | Y_0^{\delta})   ,
\label{eq: ub interms of arbit delta}
\end{eqnarray}
where the first inequality follows since $\tv_\delta$ is a refinement of  $\tv$, the equality by writing out the sum that defines $I_{\tv_\delta} (  X_0^T
\to Y_0^T )$ and isolating its first term, and the second inequality by the infimum over partitions inherent in the definition of $ I( X_0^T \to Y_\delta^T | Y_0^{\delta}) $. The arbitrariness of $\delta < t_1$ in  \eq{eq: ub interms of arbit delta} implies
\begin{equation}
\limsup_{\delta \rightarrow 0^+} I(X_0^{\delta} ; Y_0^{\delta}) + I( X_0^T \to Y_\delta^T | Y_0^{\delta}) \leq I_{\tv} (  X_0^T
\to Y_0^T )
\end{equation}
which, by the arbitrariness of  $\tv \in \Tc(0,T)$, implies
\begin{equation} \label{eq: limsup part of revision}
\limsup_{\delta \rightarrow 0^+} I(X_0^{\delta} ; Y_0^{\delta}) + I( X_0^T \to Y_\delta^T | Y_0^{\delta}) \leq I (  X_0^T
\to Y_0^T ).
\end{equation}
On the other hand, for any $\delta > 0$, we clearly have
\begin{equation} \label{eq: lb on i plus delta}
 I(X_0^{\delta} ; Y_0^{\delta}) + I( X_0^T \to Y_\delta^T | Y_0^{\delta}) \geq I (  X_0^T
\to Y_0^T ),
\end{equation}
as the right hand side, by its definition, is an infimum over all partitions in $\Tc(0,T)$, while the left hand side corresponds to an infimum over the subset consisting only of those partitions with $t_1 = \delta$. By the arbitrariness of $\delta$ in \eq{eq: lb on i plus delta} we obtain
\begin{equation}
\liminf_{\delta \rightarrow 0^+} I(X_0^{\delta} ; Y_0^{\delta}) + I( X_0^T \to Y_\delta^T | Y_0^{\delta}) \geq I (  X_0^T
\to Y_0^T )
\end{equation}
which, when combined with \eq{eq: limsup part of revision}, finally implies
\begin{equation} \label{eq: existence of the limit}
\lim_{\delta \rightarrow 0^+} I(X_0^{\delta} ; Y_0^{\delta}) + I( X_0^T \to Y_\delta^T | Y_0^{\delta}) = I (  X_0^T
\to Y_0^T ).
\end{equation}
Existence of the limit in \eq{eq: existence of the limit}, when combined with \eq{eq: conservation law in cont time} and the added assumption $ I (  X_0^T ; Y_0^T ) < \infty$, implies  existence of the limit $\lim_{\delta \rightarrow 0^+} I ( Y_0^{T-\delta} \to X_0^T ) = I ( Y_0^{T-} \to
X_0^T )$  and that
\begin{equation} \label{eq: conservation law under continuity again}
I ( X_0^T \to Y_0^T )  + I ( Y_0^{T-} \to X_0^T )
= I (  X_0^T ; Y_0^T ),
\end{equation}
thus completing the proof.
\end{IEEEproof}

\section{Directed Information, Feedback, and Causal Estimation}
\label{sec: Directed Information and Causal Estimation}
\subsection{The Gaussian Channel}
In \cite{Duncan1968}, Duncan discovered the following fundamental
relationship between the minimum mean squared error (MMSE) in causal
estimation of a target signal corrupted by an additive white Gaussian
noise (AWGN) in continuous time and the mutual information between the clean
and noise-corrupted signals:
\begin{theorem}[Duncan \cite{Duncan1968}] \label{th: Duncan68}
Let $X_0^T$ be a signal of finite average power $\int_0^T E[X_t^2]
dt < \infty$, independent of a standard Brownian motion $\{ B_t
\}$. Let $Y_0^T$ satisfy $dY_t = X_t dt + dB_t$. Then
\begin{equation}\label{eq: Duncan relation original}
\frac{1}{2}
\int_0^T E \bigl[(X_t - E[X_t | Y_0^t])^2 \bigr] dt = I (  X_0^T ; Y_0^T ).
\end{equation}
\end{theorem}
A remarkable aspect of Duncan's theorem is that the relationship
\eqref{eq: Duncan relation original} holds regardless of the
distribution of $X_0^T$. Among its ramifications is the invariance
of the causal MMSE to the flow of time,
or more generally, to any reordering of time \cite{CohenWeissmanMerhav08,
GSV2008}. It should also be mentioned that, although this exact relationship holds in continuous-time, approximate versions that hold in discrete-time can be derived from it, as is done in \cite[Theorem 9]{GSV2008}.

A key stipulation in Duncan's theorem is the independence between
the noise-free signal $X_0^T$ and the channel noise $\{ B_t \}$,
which excludes scenarios in which the evolution of $X_t$ is affected
by the channel noise, as is often the case in signal processing
(e.g., target tracking) and communication (e.g., in the presence
of feedback). Indeed, the identity \eqref{eq: Duncan relation original} does not
hold in the absence of such a stipulation.

As an extreme example, consider the case where the channel input is simply the channel output
with some delay, i.e.,
\begin{equation} \label{eq: extreme example}
X_{t+\eps} = Y_t
\end{equation}
for some $\eps > 0$ (and $X_t \equiv 0$ for $t \in
[0, \eps)$). In this case the causal MMSE on the left side of \eqref{eq: Duncan relation
original} is clearly $0$, while the mutual information on its right side is infinite. On the
other hand, in this case the directed information $I (  X_0^T \to Y_0^T ) =
0$, as can be seen by noting that $I_{\tv} (  X_0^T \to Y_0^T ) = 0$
for all $\tv$ satisfying $\max_i (t_i - t_{i-1})  \le \eps$
(since for such $\tv$, $X_0^{t_i}$ is determined by $Y_0^{t_{i-1}}$ for all $i$).

The third remark following Proposition \ref{prop: 2 properties of
directed info} implies that Theorem \ref{th: Duncan68} could be
equivalently stated with $I ( X_0^T ; Y_0^T ) $ on the right side of \eqref{eq: Duncan
relation original} replaced by $I ( X_0^T \to Y_0^T )$. Furthermore,
such a modified identity would be valid in
the extreme example in~\eqref{eq: extreme example}.
This is no coincidence and is a consequence of the
 result that follows, which generalizes Duncan's theorem. To state it formally we assume a probability space $(\Omega, \mathcal{F}, P)$ with an associated filtration $\{ \mathcal{F}_t \}$ satisfying the ``usual conditions'' (right-continuous and $\mathcal{F}_0$ contains all the $P$-negligible events in $\mathcal{F}$, cf., e.g., \cite[Definition 2.25]{KaratzasandShreve}). Recall also that when the standard Brownian motion is adapted to $\{ \mathcal{F}_t \}$ then, by definition, it is implied that, for any $s < t$, $B_t - B_s$ is independent of $ \mathcal{F}_s$ (rather than merely of $B_0^s$,  cf., e.g., \cite[Definition 1.1]{KaratzasandShreve}).
 \begin{theorem} \label{th: our generalization of Duncan's theorem}
Let $\{ (X_t, B_t)\}_{t=0}^T$ be adapted to the filtration $\{ \mathcal{F}_t \}_{t=0}^T$, where $X_0^T$  is a signal of finite average power  $\int_0^T E[X_t^2] dt < \infty$ and $B_0^T$ is a standard Brownian motion. Let     $Y_0^T$ be the output of the AWGN channel whose input is $X_0^T$ and whose noise is driven by $B_0^T$, i.e.,
\begin{equation}\label{eq: evolution of Y process in Duncan}
    dY_t = X_t dt + dB_t .
\end{equation}
Suppose that
the regularity assumptions of Proposition \ref{claim: dir inf in terms of its differential} are
satisfied for all $0 < t < T$. Then
\begin{equation}\label{eq: Duncan relation extended}
\frac{1}{2} \int_0^T E \bigl[(X_t - E[X_t | Y_0^t])^2 \bigr] dt
= I (  X_0^T \to Y_0^T ).
\end{equation}
\end{theorem}
Note that unlike in Theorem \ref{th: Duncan68},
where the channel input process is independent of the channel noise process,
in Theorem \ref{th: our generalization of Duncan's theorem} no such
stipulation exists and thus the setting in the latter accommodates the presence of
feedback.
Furthermore, since $I (  X_0^T \to Y_0^T )$
is not invariant to the direction of the flow of time in general,
Theorem \ref{th: our generalization of Duncan's theorem} implies,
as should be expected, that neither is the causal MMSE for
processes evolving in the generality afforded by the theorem.

That Theorem \ref{th: Duncan68} can be extended to accommodate the presence of feedback has been established for a communication theoretic
framework by Kadota, Zakai, and Ziv~\cite{Kadota--Zakai--Ziv1971b}.
Indeed, in  communication over the AWGN channel where $X_0^T = X_0^T(M)$ is the waveform associated with message $M$,  in the absence of feedback the Markov relation $M \to X_0^T \to Y_0^T$ implies that $I(X_0^T ; Y_0^T)$ on the right hand side of \eqref{eq: Duncan relation original}, when applying Theorem \ref{th: Duncan68} in this restricted communication framework,
can be equivalently written as  $I(M ; Y_0^T)$.
The main result of  \cite{Kadota--Zakai--Ziv1971b} is that
this relationship between the causal estimation error and $I(M ; Y_0^T)$
persists in the presence of feedback, i.e., that
\begin{equation}\label{eq: Duncan relation original again}
\frac{1}{2}
\int_0^T E \bigl[(X_t - E[X_t | Y_0^t])^2 \bigr] dt = I (  M ; Y_0^T )
\end{equation}
with or without feedback, even though, in the presence of feedback, one no longer has $I (  M ; Y_0^T ) = I (  X_0^T ; Y_0^T )$ and therefore  \eq{eq: Duncan relation original} is no longer true.
  The combination of
Theorem \ref{th: our generalization of Duncan's theorem} with   the
main result of  \cite{Kadota--Zakai--Ziv1971b} (namely, with \eq{eq: Duncan relation original again}) thus implies that in communication
over the AWGN channel, with or without feedback, we have $I(M ; Y_0^T) = I (  X_0^T \to Y_0^T )$.
This equality holds well beyond the Gaussian channel, as is elaborated in
Section \ref{sec: Communication Over Continuous-Time Channels
with Feedback}.
 Evidently, Theorem \ref{th: our generalization of Duncan's theorem} can be considered an extension of the Kadota--Zakai--Ziv result as it holds in settings  more general than communication, where
there is no message but merely a signal observed through additive
white Gaussian noise, adapted to a general filtration.

Theorem \ref{th: our generalization of Duncan's theorem} is a direct consequence of Proposition \ref{claim: dir inf in terms of its differential} and the following lemma.
\begin{lemma}[\cite{TW2010}]  \label{lemma: relative ent with fb for gaussian}
Let $P$ and $Q$ be two probability laws governing $(X_0^T, Y_0^T)$, under which \eqref{eq: evolution of Y process in Duncan} and the stipulations of Theorem \ref{th: our generalization of Duncan's theorem}  are satisfied. Then
\begin{equation}
\label{eq: main relevant result from weissman2010}
D( P_{Y_0^T} \| Q_{Y_0^T} ) = \frac{1}{2} E_P \biggl[ \int_0^T   (X_t - E_Q[X_t|Y_0^t])^2 - (X_t - E_P[X_t|Y_0^t])^2  dt  \biggr]  .
\end{equation}
\end{lemma}
Lemma \ref{lemma: relative ent with fb for gaussian} was implicit in \cite{TW2010}. It follows from
the second part of \cite[Theorem 2]{TW2010}, put together with the exposition in \cite[Subsection IV-D]{TW2010} (cf., in particular, equations  (148) through (161) therein).

\begin{IEEEproof}[Proof of Theorem \ref{th: our generalization of Duncan's theorem}]
Consider
\begin{align}
I(Y_t^{t+\delta}; X_0^{t+\delta}|Y_0^t)
&=   D ( P_{Y_t^{t+\delta} | X_t^{t+\delta}, Y_0^t } \|  P_{Y_t^{t+\delta}  |  Y_0^t } |  P_{Y_0^t, X_t^{t+\delta}}  )  \\
&=  \label{eq: integrand for rel app} \int D ( P_{Y_t^{t+\delta} | X_t^{t+\delta} = x_t^{t+\delta}, Y_0^t  = y_0^t } \|  P_{Y_t^{t+\delta}  |  Y_0^t = y_0^t }   ) d P_{Y_0^t, X_t^{t+\delta}} (y_0^t, x_t^{t+\delta}) \\
&=   \frac{1}{2}  \int  E  \biggl[  \int_t^{t+\delta}  (x_s - E[X_s |Y_0^s])^2 - (x_s - x_s)^2    ds \,\bigg|\, y_0^t, x_t^{t+\delta} \biggr] d P_{Y_0^t, X_t^{t+\delta}} (y_0^t, x_t^{t+\delta})
\label{eq: duncan proof step a} \\
&=  \frac{1}{2} \int_t^{t + \delta} E \bigl[(X_s - E[X_s | Y_0^s])^2 \bigr] ds,
\end{align}
where the equality in~\eqref{eq: duncan proof step a}
follows by applying \eqref{eq: main relevant result from weissman2010} to the integrand in \eqref{eq: integrand for rel app} as follows: replacing the time interval $[0,T)$ by $[t, t+ \delta)$, substituting $P$ by the law of $(X_t^{t+\delta}, Y_t^{t+\delta})$ conditioned on $(y_0^t, x_t^{t+\delta})$ (note that $X_t^{t+\delta}$ is deterministic at $x_t^{t+\delta}$ under this law), and substituting $Q$ by the law of $(X_t^{t+\delta}, Y_t^{t+\delta})$ conditioned on $y_0^t$. The last step is obtained by switching between the integral $\int_t^{t+\delta}$  and  $\int  E $ and then using the definition of conditional expectation. The switch between the integrals is possible due to Fubini's theorem and the fact that the signal has finite average power  $\int_0^T E[X_t^2] dt < \infty$. It follows that $i_t$ defined in \eqref{eq: it defined} exists and is given by
\begin{equation}
\label{eq: it for gaussian case}
i_t = \frac{1}{2} E \bigl[ (X_t - E[X_t | Y_0^t])^2 \bigr],
\end{equation}
which completes the proof by an appeal to Proposition \ref{claim: dir inf in terms of its differential}.
\end{IEEEproof}

\subsection{The Poisson Channel}
Consider the function $\ell : [0, \infty) \times [0, \infty) \rightarrow [0, \infty]$ given by
\begin{equation}
\ell (x, \hat{x}) = x \log (x/\hat{x}) - x +  \hat{x} .
\end{equation}
That this function is natural for quantifying the loss when estimating
nonnegative quantities is implied in \cite[Section 2]{AtarWeissman2011}, where some of its basic properties are exposed.  Among them is  that  conditional expectation is the optimal estimator not only under the
squared error loss but also under $\ell$, i.e., for any nonnegative random variable $X$ jointly distributed with $Y$,
\begin{equation}
\min_{\hat{X} (\cdot) } E \left[ \ell (X , \hat{X}(Y)) \right] = E \left[ \ell (X , E(X|Y) ) \right] ,
\end{equation}
where the minimum is over all (measurable) maps from the domain of $Y$ into $[0, \infty)$. With this loss function, the analogue  of
Duncan's theorem for the case of  doubly stochastic Poisson  process  (i.e., the intensity is a random process) can be stated as:
\begin{theorem}[{\cite{LiptserShiryaev2001, AtarWeissman2011}}] \label{th: LiptserShiryaev2001}
Let $Y_0^T$ be a doubly stochastic Poisson  process  and $X_0^T$ be its intensity process (i.e., conditioned on $X_0^T$, $Y_0^T$ is a nonhomogenous Poisson process with rate function $X_0^T$) satisfying $E \int_0^T |X_t \log X_t | dt < \infty$.
Then
\begin{equation}\label{eq: Duncan analogue relation original for Poisson}
 \int_0^T E [ \ell  (X_t , E[X_t | Y_0^t] ) ] dt
= I (  X_0^T ; Y_0^T ).
\end{equation}
\end{theorem}
We remark that for $\phi ( \alpha ) = \alpha \log \alpha$, one has
\begin{equation}
E \bigl[ \phi(X_t) - \phi ( E[X_t | Y_0^t] ) \bigr]
= E \bigl[ \ell  (X_t , E[X_t | Y_0^t] ) \bigr] ,
\end{equation}
and thus \eq{eq: Duncan analogue relation original for Poisson} can equivalently be expressed as
\begin{equation}\label{eq: Duncan analogue relation original for Poisson old}
\int_0^T E \bigl[ \phi(X_t) - \phi ( E[X_t | Y_0^t] ) \bigr] dt
= I (  X_0^T ; Y_0^T ),
\end{equation}
as was done in \cite{LiptserShiryaev2001} and other classical references. But it was not until  \cite{AtarWeissman2011} that the left hand side was established as the minimum mean causal estimation error
under an explicitly identified loss function, thus completing the analogy with Duncan's theorem.

The condition stipulated in the third item
of Proposition \ref{prop: 2 properties of directed info} is readily seen to hold when
 $Y_0^T$ is a  doubly stochastic Poisson  process and  $X_0^T$ is its intensity process.
Thus, the above theorem  could equivalently be stated with
directed information rather than mutual information on the right hand side
of \eqref{eq: Duncan analogue relation original for Poisson}.
Indeed, with continuous-time directed information replacing mutual information,
this relationship remains true in  much wider generality, as the
next theorem shows. In the statement of the theorem, we use the
notions of a point process  and its predictable intensity, as
developed in detail in, e.g.,  \cite[Chapter II]{Bremaud81}.

\begin{theorem} \label{th: our generalization of Duncan's theorem for Poisson}
Let $Y_t$ be a point process  and $X_t$ be its
$\mathcal{F}_t^Y$-predictable intensity, where $\mathcal{F}_t^Y$
is the  $\sigma$-field $\sigma (Y_0^t)$ generated by $Y_0^t$.
Suppose that  $E \int_0^T |X_t \log X_t | dt <
\infty$, and that the assumptions of Proposition \ref{claim: dir inf in terms of its differential} are satisfied for all $0 < t < T$. Then
\begin{equation}\label{eq: Duncan relation extended Poisson}
\int_0^T E [ \ell  (X_t , E[X_t | Y_0^t] ) ] dt
  = I (  X_0^T \to Y_0^T ).
\end{equation}
\end{theorem}
Paralleling the proof of Theorem \ref{th: our generalization of Duncan's theorem}, the proof of Theorem \ref{th: our generalization of Duncan's theorem for Poisson} is a direct application of Proposition \ref{claim: dir inf in terms of its differential} and the following:
\begin{lemma}[\cite{AtarWeissman2011}]  \label{lemma: relative ent with fb for poisson}
Let $P$ and $Q$ be two probability laws governing $(X_0^T, Y_0^T)$ under the setting and stipulations of Theorem \ref{th: our generalization of Duncan's theorem for Poisson}.  Then
\begin{equation}
\label{eq: main relevant result from atarweissman2011}
D( P_{Y_0^T} \| Q_{Y_0^T} ) =  E_P \left[ \int_0^T   \ell ( X_t , E_Q[X_t|Y_0^t]) -  \ell (X_t , E_P[X_t|Y_0^t])  dt  \right]  .
\end{equation}
\end{lemma}
Lemma \ref{lemma: relative ent with fb for poisson} is implicit in \cite{AtarWeissman2011}, following directly
from  \cite[Theorem 4.4]{AtarWeissman2011} and the
discussion in \cite[Subsection 7.5]{AtarWeissman2011}.
 Equipped with it, the
  proof  of Theorem \ref{th: our generalization of Duncan's theorem for Poisson}  follows similarly as that of  Theorem \ref{th: our generalization of Duncan's theorem},  the role of \eqref{eq: main relevant result from weissman2010} being played here by \eq{eq: main relevant result from atarweissman2011}.

\section{Example: Poisson Channel with Feedback \label{sec: poisson feedback example}}

The Poisson channel (e.g., \cite{Mazo76PoissonChannelIntro,
kabanov78CapacityPoissonChannel, Davis80PoissonCapacity,
Wyner88_Possion_channel, Wyner88_Possion_channel2, Lapidoth93_Poisson,
GuoShamaiVerdue08Poisson, BrossLapidoth09Poisson}) is a channel where the input  at time $t$,  $X_t$, determines the intensity of the doubly stochastic Poisson process $Y_t$ occurring at the output of the channel. A Poisson channel with feedback refers to the case where the input signal $X_t$ may depend on  the previous observation of the output $Y^{t}$.

In this section we consider a special case of Poisson channel with feedback. Let $\mathbf{X} = \{X_t\}$ and $\mathbf{Y} = \{Y_t\}$ be the input and
output processes of the continuous-time Poisson channel with feedback,
where each time an event occurs at the channel output, the channel input
changes to a new value, drawn according to the distribution of a
positive random variable $X$, independently of the channel input and
output up to that point in time. The channel input remains fixed at that
value until the occurrence of the next event at the channel output, and
so on.  Throughout this section, the shorthand  ``Poisson channel with feedback'' will refer to this scenario, with its implied channel input process.

The Poisson channel we use here is similar to the well-known
Poisson channel model (e.g., \cite{Mazo76PoissonChannelIntro,
kabanov78CapacityPoissonChannel, Davis80PoissonCapacity,
Wyner88_Possion_channel, Wyner88_Possion_channel2, Lapidoth93_Poisson,
GuoShamaiVerdue08Poisson, BrossLapidoth09Poisson}) with one difference
that the intensity of the Poisson channel changes according to the input
$X$ only when there is an event at the output of the channel. Note that
the channel description given here uniquely determines the
joint distribution of the channel input and output processes.

In the first part of this section, we derive, using Theorem \ref{th:
our generalization of Duncan's theorem for Poisson}, a formula for the
directed information rate of this Poisson channel with feedback. In the
second part, we demonstrate the use of this formula by  computing and plotting the directed information
rate for a special case in which the intensity alphabet is of size 2.

\subsection{Characterization of the Directed Information Rate}
For jointly distributed processes $(\mathbf{X},  \mathbf{Y})$ define the directed information rate $I( \mathbf{X} \to \mathbf{Y})$ by
\begin{equation} \label{eq: directed info rate def}
I( \mathbf{X} \to \mathbf{Y})
= \lim_{T \to \infty} \frac{1}{T} I (X_0^T \to Y_0^T),
\end{equation}
when the limit exists.

\begin{proposition} \label{proposition: directed info in poisson channel example}
Assume that $X$ is
finite-valued with probability mass function (pmf) $p_X(x)$. The directed information rate between the input and output processes of  the Poisson channel with feedback $I( \mathbf{X} \to \mathbf{Y})$
exists and is given by
\begin{equation} \label{eq: directed info expression for our poisson example}
I( \mathbf{X} \to \mathbf{Y})
= \frac{I(X;Y)}{E[1/X]},
\end{equation}
where, in $I(X;Y)$ on the right hand side, $Y | \{X = x\} \sim \mathrm{Exp}(x)$,
i.e., the conditional density of\, $Y$ given $\{X=x\}$ is
$f(y|x)=xe^{-yx} \cdot 1_{\{ y \geq 0 \}}$.
\end{proposition}
The key component in the proof of the proposition is the use of
Theorem~\ref{th: our generalization of Duncan's theorem for Poisson} for
directed information in continuous time as a causal mean estimation
error.  An intuition for the expression in
\eq{eq: directed info expression for our poisson example} can be
obtained by considering rate per unit cost \cite{Verdu90}, i.e., $R =
I(X;Y)/E[b(X)]$, where $b(x)$ is the cost of the input. In our case, the
``cost'' of $X$ is proportional to the average duration of time until
the channel can be used again, i.e.,  $b(x)= 1/x$. Finally, we remark that the assumption of  discreteness of $X$ in Proposition
\ref{proposition: directed info in poisson channel example} is  made for simplicity of the proof, though the result carries over to more generally distributed $X$.

To prove Proposition~\ref{proposition: directed info in poisson channel example}, let us first collect the following observations:

\begin{lemma} \label{lemma: observations collection for our prop}
Let $X \sim p_X(x)$ and $Y | \{X = x\} \sim \mathrm{Exp}(x)$. Define
\begin{equation}
\label{eq: conditional expectation as func of t g}
g(t) := E[X | Y \ge t] = \frac{\sum_{x} x e^{-t x} p_X(x)}{\sum_{x} e^{-t x} p_X(x)}, \quad t \ge 0.
\end{equation}
Then the following statement holds.
\begin{enumerate}
\item The marginal distribution of $X_t$ is
\begin{equation} \label{eq: pmf of Xt}
P\{X_t = x\} = \frac{(1/x) p_X(x)}{\sum_{x'} (1/x') p_X(x')}
\end{equation}
and consequently
 \begin{equation} \label{eq: exp of xtlogxt}
E[X_t \log X_t ] = \frac{E [\log X]}{E [1/X]}.
\end{equation}

\item Let   $\ell = \ell (Y_{-\infty}^0)$ denote the time of occurrence of the last (most recent) event at the channel output prior to time $0$ and define   $\tau := -\ell$. The density of $\tau$ is
\begin{equation} \label{eq: density of tau}
  f_{\tau} (t) = \frac{\sum_x e^{-t x} p_X(x)}{E [ 1/X ]} , \quad t \ge 0.
\end{equation}

\item For $\tau$ distributed as in \eqref{eq: density of tau},
\begin{equation} \label{eq: expectation of g log g of t}
 E [g(\tau) \log g (\tau)] = \frac{1 - h(Y)}{E[1/X]}.
\end{equation}
\end{enumerate}
\end{lemma}

\begin{IEEEproof}
For the first part of the lemma,
note that $X_t$ is an ergodic continuous-time Markov chain
and thus $P\{X_t = x\}$ is equal to
the fraction of time  that $X_t$ spends in state $x$ which is proportional to
 $(1/x) p_X(x)$,
accounting for \eqref{eq: pmf of Xt}, which, in turn, yields
\begin{equation}
E[X_t \log X_t ]
= \sum_x \frac{(1/x)p_X(x)}{\sum_{x'} (1/x')p_X(x')} x \log x
= \frac{ \sum_x p_X(x)  \log x}{\sum_{x'} (1/x') p_X(x')}
= \frac{E [\log X]}{E [1/X]},
\end{equation}
accounting for \eqref{eq: exp of xtlogxt}.

To prove the second part of the lemma, observe that
\begin{enumerate}
\item[(a)]
the interarrival times of the process $\mathbf{Y}$ are
independent and identically distributed (i.i.d.) copies
of a random variable $Y$;
\item[(b)] $Y$ has a density
\begin{equation} \label{eq: density of Y}
f_Y(y) = \sum_x p_X(x) x e^{-x y}, \quad y \ge 0,
\end{equation}
\item[(c)] 
the probability density of the length of the interarrival interval of the $\mathbf{Y}$ process around $0$ is proportional to $f_Y(y) \cdot y$; and
\item[(d)] 
given the length of the interarrival interval around $0$ is $y$, its left point is uniformly
distributed on $[-y, 0]$.
\end{enumerate}
Letting $\mbox{Unif}[0,y]( \cdot) $ denote the density of a random variable uniformly distributed on $[0,y]$, it  follows that the density of $\tau$ is
\begin{align}
f_{\tau} (t)
&= \label{eq: f tau step a}
\int_0^\infty \frac{f_Y(y) \cdot y}{\int_0^\infty f_Y(y') \cdot y' dy'} \mbox{Unif}[0,y](t)   dy \\
&= \int_t^\infty \frac{f_Y(y) \cdot y}{\int_0^\infty f_Y(y') \cdot y' dy'} \frac{1}{y}   dy \\
&= \label{eq: f tau step c}
\frac{  \sum_x p_X(x) x \int_t^\infty e^{-x y} dy }{ \sum_x p_X(x) x \int_0^\infty e^{-x y'} \cdot y' dy'} \\
&= \frac{  \sum_x p_X(x) x \frac{e^{-tx}}{x}}{ \sum_x p_X(x) x \frac{1}{x^2}} \\
&= \frac{  \sum_x p_X(x)  e^{-tx}}{ E[1/X]},
\end{align}
where \eqref{eq: f tau step a} follows by combining observations (c) and (d), and \eqref{eq: f tau step c}
follows by substituting from \eqref{eq: density of Y}. We have thus proven the second part of the lemma.

To establish the third part, let $F_Y (t)$
denote the cumulative distribution function of $Y$ and consider
\begin{align}
E [g(\tau) \log g (\tau)]
&=   \int_0^\infty  f_{\tau} (t) g(t) \log g(t) \\
&= \label{eq: cdf of y step a}
\int_0^\infty \frac{  \sum_x p_X(x)  e^{-tx}}{ E[1/X]} \frac{\sum_{x} x e^{-t x} p_X(x)}{\sum_{x} e^{-t x} p_X(x)} \log \frac{\sum_{x} x e^{-t x} p_X(x)}{\sum_{x} e^{-t x} p_X(x)} dt  \\
&= \frac{1}{E[1/X]}
\int_0^\infty \sum_{x} x e^{-t x} p_X(x) \log \frac{\sum_{x} x e^{-t x} p_X(x)}{\sum_{x} e^{-t x} p_X(x)} dt   \\
&= \label{eq: cdf of y step b}
\frac{1}{E[1/X]} \int_0^\infty f_Y(t) \log \frac{f_Y(t)}{1-F_Y(t)} dt \\ &= \frac{1}{E[1/X]} \biggl( \int_0^\infty f_Y(t) \log \frac{1}{1-F_Y(t)} dt  - h(Y) \biggr) \\
&=  \frac{1}{E[1/X]} \biggl( \int_0^1  \log \frac{1}{1- u} du  - h(Y) \biggr) \\
&= \frac{1}{E[1/X]} ( 1  - h(Y) ),
\end{align}
where \eqref{eq: cdf of y step a} follows by substituting from the second part of the lemma and \eqref{eq: cdf of y step b} follows by substituting from \eqref{eq: density of Y} and noting that
\begin{align}
\sum_{x} e^{-t x} p_X(x)
= \sum_x p_X(x) x \frac{e^{-tx}}{x}
&= \sum_x p_X(x) x \int_t^\infty e^{-x y} dy \nonumber \\
&= \int_t^\infty \sum_x p_X(x) x e^{-x y} dy
= \int_t^\infty f_Y(y) dy
= 1-F_Y(t).
\end{align}
We have thus established the third and last part of the lemma.
\end{IEEEproof}

\begin{IEEEproof}[Proof of Proposition \ref{proposition: directed info in poisson channel example}]
We have
\begin{align}
I( \mathbf{X} \to \mathbf{Y})
&= \lim_{T \to \infty} \frac{1}{T} I (X_0^T \to
    Y_0^T) \\
&= \label{eq: poisson step a}
\lim_{T \to \infty} \frac{1}{T} \int_0^T E \bigl[ X_t \log X_t - E[X_t | Y_0^t] \log E[X_t | Y_0^t] \bigr] dt  \\
&= \label{eq: poisson step b}
E \bigl[ X_0 \log X_0 - E[X_0 | Y_{-\infty}^0] \log E[X_0 | Y_{-\infty}^0] \bigr]  \\
&=
\frac{E [\log X]}{E [1/X]} -  E \bigl[  E[X_0 | Y_{-\infty}^0] \log E[X_0 | Y_{-\infty}^0] \bigr],
\label{eq: first part eq in pf of prop}
\end{align}
where \eqref{eq: poisson step a} follows from the relation between directed information and causal estimation
in \eqref{eq: Duncan relation extended Poisson};
\eqref{eq: poisson step b} follows from the stationarity and martingale convergence. Specifically, by martingale convergence
$E[X_0| Y_{-t}^0] \rightarrow E[X_0| Y_{-\infty}^0]$ as $t \rightarrow \infty$ a.s.\ and thus $E \bigl[ X_t \log X_t - E[X_t | Y_0^t] \log E[X_t | Y_0^t] \bigr]$, which by stationarity is equal to  $ E \bigl[ X_0 \log X_0 - E[X_0| Y_{-t}^0] \log E[X_0| Y_{-t}^0] \bigr] $, converges to
$E \bigl[ X_0 \log X_0 - E[X_0 | Y_{-\infty}^0] \log E[X_0 | Y_{-\infty}^0] \bigr] $ by the bounded convergence theorem (recall that $X_0$ is finite-valued);
and \eqref{eq: first part eq in pf of prop} follows from the first part of Lemma \ref{lemma: observations collection for our prop}. Now,
recalling the definition of the function $g$ in \eqref{eq: conditional expectation as func of t g}
we note that
\begin{equation} \label{eq: conditional exp interms of g}
E[X_0 | \ell (Y_{-\infty}^0) ] = g ( - \ell (Y_{-\infty}^0) ).
\end{equation}
Thus
\begin{align}
E \bigl[  E[X_0 | Y_{-\infty}^0] \log E[X_0 | Y_{-\infty}^0] \bigr]
&= \label{eq: E with g step a}
E \bigl[ E[X_0 | \ell (Y_{-\infty}^0) ] \log E[X_0 | \ell (Y_{-\infty}^0) ] \bigr] \\
&= \label{eq: E with g step b}
E \bigl[ g ( - \ell (Y_{-\infty}^0) ) \log g ( - \ell (Y_{-\infty}^0)) \bigr]  \\
&= E [ g ( \tau ) \log g ( \tau ) ] \\
&= \frac{1 - h(Y)}{E[1/X]}, \label{eq: 2nd part eq in prop}
\end{align}
where \eqref{eq: E with g step a} follows from the Markov relation $Y_{-\infty}^0 \to  \ell (Y_{-\infty}^0) \to X_0$, \eqref{eq: E with g step b} follows from \eqref{eq: conditional exp interms of g}, and \eqref{eq: 2nd part eq in prop} from the last part of Lemma \ref{lemma: observations collection for our prop}.
Thus
\begin{align}
I( \mathbf{X} \to \mathbf{Y})
&= \label{eq: poisson proof last step a}
\frac{h(Y) -1 + E [\log X]}{E [1/X]} \\
&= \label{eq: poisson proof last step b}
\frac{h(Y) - h(Y|X)}{E [1/X]} \\
&= \frac{I(X;Y)}{E [ 1/X ]},
\end{align}
where \eqref{eq: poisson proof last step a} follows by  combining \eqref{eq: first part eq in pf of prop} with \eqref{eq: 2nd part eq in prop}, and \eqref{eq: poisson proof last step b} follows by noting that
\begin{equation}
h(Y|X) = \sum_x h(Y|X=x) p_X(x) = \sum_x (1 - \log x ) p_X(x) = 1 - E[\log X].
\end{equation}
This completes the proof of Proposition~\ref{proposition: directed info in poisson channel example}.
\end{IEEEproof}

\subsection{Evaluation of the Directed Information Rate}

\begin{figure}[h!]{
\psfrag{0}[][][1]{} \psfrag{0.4}[][][1]{} \psfrag{0.8}[][][1]{} \psfrag{0.04}[][][1]{}
\psfrag{0.08}[][][1]{}\psfrag{0.12}[][][1]{} \psfrag{p}[][][1]{$p := P\{X=\lambda_1\}$}
\psfrag{0.4}[][][1]{} \psfrag{0.8}[][][1]{} \psfrag{direc}[][][1]{$I( \mathbf{X} \to
\mathbf{Y})$} \psfrag{M}[][][1]{}

\centerline{
\includegraphics[width=7cm]{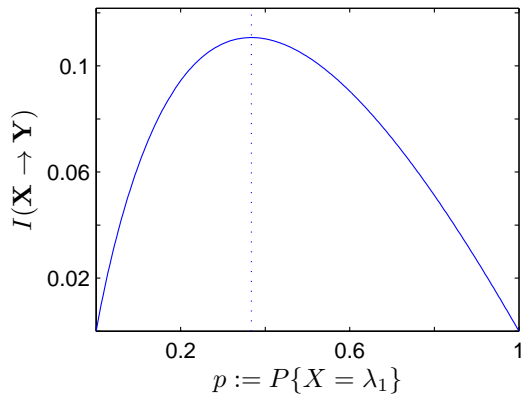}}
\caption{The directed information rate between the input and output processes for the continuous-time Poisson channel with feedback, as a function of
$P(x)$, the pmf of the input to the channel. The input to the channel is one of two possible values
$\lambda_1=1$ and $\lambda_2=2$, and it is the intensity of the Poisson process at the output of
the channel until the next event. }\label{f_poisson1} }\end{figure}

Fig.~\ref{f_poisson1} depicts the  directed information rate $I(\mathbf{X} \to \mathbf{Y})$ for the case where $X$ takes only two values
$\lambda_1$ and $\lambda_2$. We have used  numerical evaluation of
$I(X;Y)$ in the right hand side of \eqref{eq: directed info expression for our poisson example}  to compute the directed information rate. The figure  shows the influence of $p = P\{X=\lambda_1\}$ on the directed information rate where $\lambda_1=1$ and $\lambda_2=2$. As
expected, the maximum is achieved when there is higher probability that the
encoder output will be the higher rate $\lambda_2$, which would imply more channel uses per unit time,  but not much higher as otherwise the input value will be close to deterministic.

\begin{figure}[h!]{
\psfrag{lambda}[][][1]{$\lambda_2$} \psfrag{direc}[][][1]{$\max I(\mathcal X \to \mathcal Y)$}
\psfrag{M}[][][1]{} \centerline{
\includegraphics[width=7cm]{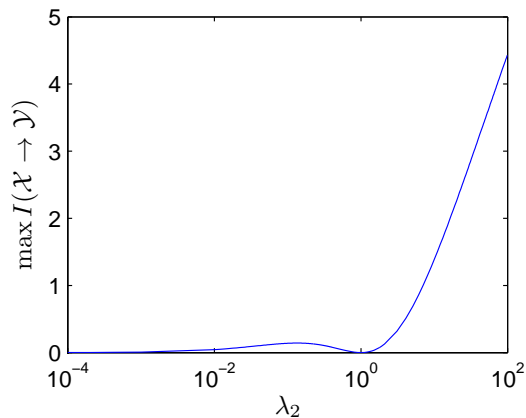}}
\caption{Capacity of the Poisson channel with feedback, in case where channel input is constrained to the binary set $\{ \lambda_1, \lambda_2 \}$,  when $\lambda_1$ is fixed and is equal to
1 and $\lambda_2$ varies.}\label{f_poisson_lambda2} }\end{figure}

Fig. \ref{f_poisson_lambda2} depicts the maximal value (optimized w.r.t.\  $P \{ X = \lambda_1
\}$) of the directed information rate  when $\lambda_1$ is fixed and is equal to 1 and
$\lambda_2$ varies. 
This value is the
capacity of the Poisson channel  with feedback,  when the inputs are restricted to one of the
two values $\lambda_1$ or $\lambda_2$.
When $\lambda_2=0$  the capacity  is obviously zero since any
use of $X=\lambda_2$ as input will cause the channel not to change any further. It is also obviously zero at
$\lambda_2=1$ since in this case $\lambda_1=\lambda_2$, so there is only one possible input to the channel.  As $\lambda_2$ increases, the capacity of
the channel increases without bound since, for $\lambda_2 \gg \lambda_1$, the channel effectively operates as a noise-free binary channel, where one symbol ``costs'' an average duration of $1$ while the other a vanishing average duration. Thus the limiting capacity with increasing $\lambda_2$
is equal to $\lim_{p \downarrow 0} H(p)/p = \infty$.


One can consider a discrete-time memoryless channel,
where the input $X$ is discrete ($\lambda_1$ or
$\lambda_2$) and the output $Y$ is distributed according to $\mathrm{Exp}(X)$.
Consider now a random cost $b(X)=Y$, where $Y$ is the output of the channel.
Using the result from \cite{Verdu90} we obtain that the capacity per
unit cost of the discreet memoryless channel is
\begin{equation}
\max_{P(x)}\frac{I(X;Y)}{E[Y]}
=\max_{P(x)}\frac{I(X;Y)}{E[1/X]},
\end{equation}
where the equality follows
since $E[Y]=E[E[Y|X]]=E[1/X]$.
Finally, we note that the capacity of the Poisson channel in the example
above is the
capacity per unit cost of the discrete memoryless channel. Thus,
by Proposition \ref{proposition: directed info in poisson channel example}
we can conclude that the continuous-time directed information rate
characterizes the capacity of the Poisson channel with feedback.
In the next section we will see that the continuous-time directed information
rate characterizes the capacity of a large family of continuous-time
channels.

\section{Communication over Continuous-Time Channels
with Feedback} \label{sec: Communication Over Continuous-Time Channels with Feedback}

We first review the definition of a block-ergodic process as given by Berger
\cite{Berger68_blockergodic}. Let $(X,\mathcal{X},\mu)$ denote a continuous-time process
$\{X_t\}_{t \ge 0}$ drawn from a space $\mathcal{X}$ according to the probability measure $\mu$.
For $t > 0$, let $T^t$ be a $t$-shift transformation, i.e., $(T^tx)_s=x_{s+t}$. A measurable set
$\mathcal A$ is {\it $t$-invariant} if it does not change under the $t$-shift transformation,
i.e., $T^t\mathcal A=\mathcal A$. A continuous-time process $(X,\mathcal X, \mu)$ is {\it
$\tau$-ergodic} if every measurable $\tau$-invariant set of processes has either probability 1 or
0, i.e.,  for any $\tau$-invariant set $\mathcal{A}$, in other words,
$\mu(\mathcal{A})=(\mu(\mathcal{A}))^2$. The definition of $\tau$-ergodicity means that if we
take the process $\{X_t\}_{t\ge 0}$ and slice it into time-blocks of length $\tau$, then the new
discrete-time process $(X_0^{\tau},X_\tau^{2\tau},X_{2\tau}^{3\tau},\ldots)$ is ergodic. A
continuous-time process $(X,\mathcal X, \mu)$ is {\it block-ergodic} if it is $\tau$-ergodic for
every $\tau>0$. Berger \cite{Berger68_blockergodic} showed that weak mixing (therefore also
strong mixing) implies block ergodicity.

\begin{figure}[h!]{
\psfrag{A}[][][1]{} \psfrag{m3}[][][1]{Message\;\;}
\psfrag{m4}[][][1]{\;\;\;\;\;\;\;\;\;\;\;\;\;\;\;\;\;\; Message estimate}
\psfrag{m1}[][][1]{$M\in\{1,\ldots,2^{nT}\}\;\;\;\;\;\;$} \psfrag{X}[][][1]{$X_t$}
\psfrag{De}[][][1]{Delay $\Delta$} \psfrag{Y2}[][][1]{$\;\;\;\; Y_{t-\Delta}$}
\psfrag{Y}[][][1]{$Y_t$}
\psfrag{Y1}[][][1]{$\hat M$} 
\psfrag{D}[][][1]{$x_t (m, y_0^{t-\Delta})$} \psfrag{T1}[][][1]{Encoder} \psfrag{P}[][][1]{$g
(X_t, Z_t)$}\psfrag{T2}[][][1]{Channel} \psfrag{W}[][][1]{$\hat
m(y_0^T)$}\psfrag{T3}[][][1]{Decoder}
\centerline{
\includegraphics[width=11cm]{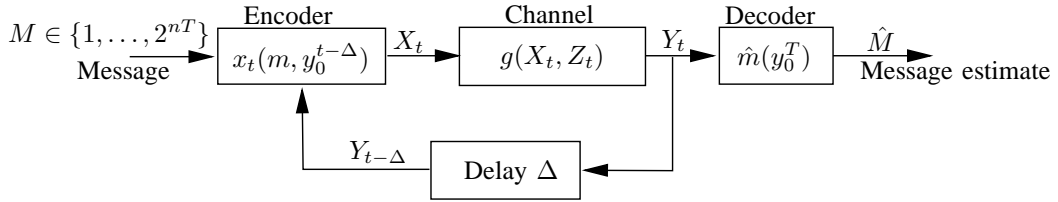}}
\caption{Continuous-time communication with delay $\Delta$ and channel of the form $Y_t =  g(X_t, Z_t),$ where $Z_t$ is a block ergodic process.}\label{f_channel_con} }\end{figure}

Now let us describe the communication model of our interest (see Fig. \ref{f_channel_con}) and
show that the continuous-time directed information characterizes the capacity. Consider a
continuous-time channel that is specified by
\begin{itemize}
\item the channel input and output alphabets
$\mathcal{X}$ and $\mathcal{Y}$, respectively, that are not necessarily finite, and
\item the channel output at time $t$
\begin{equation} \label{eq: channel output process}
Y_t =  g(X_t, Z_t)
\end{equation}
corresponding to the channel input $X_t$ at time $t$, where $\{ Z_t \}$ is a stationary ergodic
noise process on an alphabet $\mathcal{Z}$ and $g: \mathcal{X} \times \mathcal{Z} \to
\mathcal{Y}$ is a given measurable function.
\end{itemize}
A $(2^{TR}, T)$ code with delay $\Delta > 0$ for the channel consists of
\begin{itemize}
\item a message set $\{1,2,\ldots, 2^{\lfloor T R \rfloor}  \}$,
\item an encoder that assigns a symbol
\begin{equation} \label{e_encoding}
x_t(m, y_0^{t-\Delta})
\end{equation}
to each message $m \in \{1,2,\ldots,  2^{\lfloor T R \rfloor}  \}$ and past received output signal
$y_0^{t-\Delta} \in \mathcal{Y}^{[0,t-\Delta)}$ for $t \in [0,T)$,
where $x_t: \{1,2,\ldots, 2^{\lfloor T R \rfloor}  \} \times \mathcal{Y}^{[0,t-\Delta)} \to \mathcal{X}$ is measurable,
and
\item a decoder that assigns a message estimate $\hat{m}(y_0^T) \in \{1,2, \ldots, 2^{\lfloor T R \rfloor} \}$
to each received output signal $y_0^T \in \mathcal{Y}^{[0,T)}$, where $\hat{m}: \mathcal{Y}^{[0,T)} \to \{1,2, \ldots, 2^{\lfloor T R \rfloor} \}$ is measurable.
\end{itemize}
We assume that the message $M$ is uniformly distributed on $\{1,2,\ldots, \lfloor 2^{T R}
\rfloor\}$ and independent of the noise process $\{Z_t\}$.

By the definition of the channel in \eqref{eq: channel output
process}, the definition of the encoding function in (\ref{e_encoding}), and the independence of $M$ and $\{Z_t\}$, it follows that
for any $\delta > 0$ and any $t \ge 0$,
\begin{equation}M \to (X_0^{t+\delta}, Y_0^t ) \to Y_t^{t+\delta}\label{e_cdfy_Markov} \end{equation}
form a Markov chain. This is analogous to the assumption in the discrete case that $p(y_{n+1}|x^{n+1},y^n,m)=
p(y_{n+1}|x^{n+1},y^n)$; the analogy is exact when we convert a discrete time channel to a continuous time channel with constant piecewise process between the time samples.
Furthermore, for any $t \ge 0$, $\delta > 0$, and $\Delta \ge \delta$,
\begin{equation}\label{e_cdfx_Markov}
X_t^{t+\delta} \to (X_0^t, Y_0^{t+\delta - \Delta} ) \to Y_{t + \delta - \Delta}^{t + \delta}
\end{equation}
form a Markov chian. This is analogous to the assumption in the discrete case that whenever there is feedback of delay
$d\ge 1$, $p(x_{n+1}|x^n,y^n)=p(x_{n+1}|x^n,y^{n+1-d})$.

%


Similar communication settings with feedback in continuous time were studied by Kadota, Zakai, and Ziv
\cite{KadotaZakaiZiv71MemorylessContinuous} for continuous-time memoryless channels, where it is shown that
feedback does not increase the capacity, and by Ihara~\cite{Ihara94ContinuousGaussianFeedback,
IharaBook} for the Gaussian case. Our main result in this section is showing that the operational
capacity, defined below, can be characterized by the information capacity, which is the maximum
of directed information from the channel input process to the output process. Next we define an
achievable rate, the operational feedback capacity, and the information feedback capacity for our
setting.

\begin{definition} A
rate $R$ is said to be \emph{achievable with feedback delay $\Delta$} if for each $T$ there
exists a family of $(2^{RT},T)$ codes such that
\begin{equation}\label{eq: towrds rdelta acheivability}
   \lim_{T \to
    \infty} P\{M \ne \hat{M} (Y_0^T)\} = 0.
\end{equation}
\end{definition}

\begin{definition} Let
\begin{equation}\label{eq: delta delay capacity}
    C(\Delta) = \sup \{ R : R \mbox{ is achievable with feedback delay $\Delta$} \}
\end{equation}
be the {\it (operational) feedback capacity} with delay $\Delta$, and let the {\it
(operational) feedback capacity} be
\begin{equation}
C\triangleq \sup_{\Delta>0}C(\Delta).
\end{equation}
\end{definition}
From the monotonicity of $C(\Delta)$ in $\Delta$ we have $\sup_{\Delta>0}C(\Delta)=\lim_{\Delta
\to 0}C(\Delta)$. This definition coincides with the feedback capacity definition of continuous time channels given in \cite{KadotaZakaiZiv71MemorylessContinuous}, where there also was assumed a positive but arbitrary small delay in the feedback capacity.

\begin{definition} Let $C^I(\Delta)$ be the information feedback capacity defined as
\begin{equation} \label{eq: informational characterization of capacitya}
C^{I}({\Delta}) = \lim_{T \to \infty}\ \frac{1}{T}  \sup_{\mathcal{S}_\Delta} I (  X_0^T \to
Y_0^T ),
\end{equation}
where  the supremum in \eqref{eq: informational characterization of capacitya} is over
$\mathcal{S}_\Delta$, which is the set of all channel input processes of the form
\begin{equation} \label{e_xt_form}
X_t = \begin{cases}
g_t (U_t, Y_0^{t-\Delta})& t\ge \Delta,\\
g_t (U_t) & t<\Delta,
\end{cases}
\end{equation}
some family of measurable functions $\{ g_t \}_{t=0}^T$, and some process $U_0^T$ which is independent of
the channel noise process $Z_0^T$ (appearing in \eqref{eq: channel output process}) and has a
finite cardinality that may depend on $T$.
\end{definition}

The limit in \eqref{eq: informational characterization of capacitya} is shown to exist in  Lemma
\ref{l_superadditivty} using the superadditivity property.  We now characterize $C(\Delta)$ in
terms of $C^I(\Delta)$ for the class of channels defined in  \eqref{eq: channel output process}.

\begin{theorem}\label{t_capacity}
For the channel defined in \eqref{eq: channel output process},
\begin{align} \label{e_ach_delta}
C(\Delta) &\le  C^{I}(\Delta),\\
C(\Delta) &\ge  C^{I}(\Delta') \quad\text{for all } \Delta'>\Delta. \label{e_cdelta'}
\end{align}
\end{theorem}

Since $C^I(\Delta)$ is a decreasing function in  $\Delta$, \eqref{e_cdelta'} may be written as
$C(\Delta) \ge  \lim_{\delta \to \Delta^+} C^I(\delta)$, and the limit exists because of the
monotonicity. Since the function is monotonic then  $C^I(\Delta)=\lim_{\delta \to \Delta^+}
C^I(\delta)$ with a possible exception of the points of $\Delta$ of a set of measure zero
\cite[p. 5]{Riesz_90_functional_analysis}. Therefore $C(\Delta) = C^I(\Delta)$ for any $\Delta\ge
0 $ except of a set of points of measure zero. Furthermore \eqref{e_ach_delta} and
\eqref{e_cdelta'} imply that $\sup_{\Delta>0} C(\Delta)=\sup_{\Delta>0} C^{I}(\Delta)$, hence we
also have $C= \sup_{\Delta>0}C^{I}(\Delta)=\lim_{\Delta\to 0}C^{I}(\Delta)$.

Before proving the theorem we show that the limits in \eqref{eq: informational characterization
of capacitya} exist.

\begin{lemma}\label{l_superadditivty}
The term $\sup_{\mathcal{S}_\Delta} I (  X_0^T \to Y_0^T)$ is superadditive, namely,
\begin{equation}\label{e_lemma1}
\sup_{\mathcal{S}_\Delta} I (  X_0^{T_1+T_2} \to Y_0^{T_1+T_2})\ge
\sup_{\mathcal{S}_\Delta} I (  X_0^{T_1} \to Y_0^{T_1})+\sup_{\mathcal{S}_\Delta} I
( X_0^{T_2} \to Y_0^{T_2}),
\end{equation}
and therefore  the limit in \eqref{eq: informational characterization of capacitya} exists and is
equal to
 \begin{equation}\label{e_lemma2}
  \lim_{T \to \infty} \frac{1}{T}  \sup_{\mathcal{S}_\Delta}
I (  X_0^T \to Y_0^T
  )= \sup_T \frac{1}{T}  \sup_{\mathcal{S}_\Delta}
I (  X_0^T \to Y_0^T
  )
  \end{equation}
\end{lemma}

To prove Lemma \ref{l_superadditivty} we use the following result:
\begin{lemma}\label{l_di_inequality}
Let $\{(X_i,Y_i)\}_{i=1}^{n+m}$ be a pair of discrete-time processes such that Markov relation $X_i \to
(X^{i-1},Y^{i-1}) \to (X_{n+1}^{i-1},Y_{n+1}^{i-1})$ holds for $i\in\{n+1,n+2,\ldots,n+m\}$. Then
\begin{equation} \label{e_claim2}
I (  X^{n+m} \to Y^{n+m})\ge I (  X^{n} \to Y^{n})+ I (  X_{n+1}^{n+m}
\to Y_{n+1}^{n+m}),\end{equation}
\end{lemma}

\begin{IEEEproof}
The result is a consequence of the identity \cite[Eq.
(11)]{Kim07_feedback}
\begin{equation}\label{e_kim_identuty}
I(X^n\to Y^n)=\sum_{i=1}^n I(X_i;Y_{i}^n|X^{i-1},Y^{i-1}).
\end{equation}

Consider
\begin{align}
I (  X^{n+m} \to
Y^{n+m})
&=
\sum_{i=1}^{n+m}I(X_i;Y_{i}^{n+m}|X^{i-1},Y^{i-1})\label{e_1}\\
&=
\sum_{i=1}^{n}I(X_i;Y_{i}^{n+m}|X^{i-1},Y^{i-1})+\sum_{i=n+1}^{n+m}I(X_i;Y_{i}^{n+m}|X^{i-1},Y^{i-1})\\
&\geq
\sum_{i=1}^{n}I(X_i;Y_{i}^n|X^{i-1},Y^{i-1})+\sum_{i=n+1}^{n+m}I(X_i;Y_{i}^{n+m}|X_{n+1}^{i-1},Y_{n+1}^{i-1})\label{e_2}\\
&=
I (  X^{n} \to Y^{n})+ I (  X_{n+1}^{n+m} \to Y_{n+1}^{n+m}),
\end{align}
where \eqref{e_1} follows from the identity given in \eqref{e_kim_identuty}, and \eqref{e_2}
follows from the Markov chain assumption in the lemma.
\end{IEEEproof}

\begin{IEEEproof}[Proof of Lemma \ref{l_superadditivty}]
First note that we do not increase the term $\inf_\tv I_\tv (
X_0^{T_1+T_2} \to Y_0^{T_1+T_2})$ by restricting the time-partition $\tv$ to have an
interval starting at point $T_1$. Now fix three time-partitions: $\tv_1$ in $[0, T_1)$,
$\tv_2$ in $[T_1, T_1+T_2)$, and $\tv$ in $[0,T_1+T_2)$  such that $\tv$ is a
concatenation $\tv_1$ and $\tv_2$. For $X_0^{T_1}$ and $X_{T_1}^{T_1+T_2}$, fix the
input functions of the form of \eqref{e_xt_form} and fix the arguments $U^{T_1}$ and
$U_{T_1}^{T_1+T_2}$ which corresponds to $X_0^{T_1}$ and $X_{T_1}^{T_1+T_2}$, respectively. The
construction is such that the random processes $U^{T_1}$ and $U_{T_1}^{T_1+T_2}$ are independent
of each other. Let $X_0^{T_1+T_2}$ be a concatenation of $X_0^{T_1}$ and $X_{T_1}^{T_1+T_2}$.
Applying Lemma \ref{l_di_inequality} on the discrete-time process $\{(X_i,Y_i)\}_{i=1}^{n+m}$,
where $(X_i,Y_i)=(X_{t_i}^{ti+1},Y_{t_i}^{t_{i+1}})$ for $i=1,2,\ldots,n+m$  we obtain that for any
fixed $\tv_1$, $\tv_2$, $X_0^{T_1}$, $X_{T_1}^{T_1+T_2}$, $U^{T_1}$, and
$U_{T_1}^{T_1+T_2}$ as described above, we have
\begin{equation} \label{e_claim}
I_{\tv} (  X_0^{T_1+T_2} \to Y_0^{T_1+T_2})\ge I_{\tv_1} (
X_0^{T_1} \to Y_0^{T_1})+ I_{\tv_2} (  X_{T_1}^{T_1+T_2} \to
Y_{T_1}^{T_1+T_2}).
\end{equation}
Note that the Markov condition $X_i \to (X_0^{i-1},Y^{i-1}) \to (X_{n+1}^{i-1},Y_{n+1}^{i-1})$
indeed holds  because of the construction of $X_0^{T_1+T_2}$. Furthermore, because of the
stationarity
of the noise \eqref{e_claim} implies \eqref{e_lemma1}. 
Finally, using Fekete's lemma \cite[Ch.~2.6]{CombinatorialOptimization_SchrijverBook} and the
superadditivity in \eqref{e_lemma1} implies the existence of the limit in \eqref{e_lemma2}.
\end{IEEEproof}

 The proof of
Theorem~\ref{t_capacity} consists of two parts: the proof of the converse, i.e.,
\eqref{e_ach_delta}, and the proof of achievability, i.e.,  \eqref{e_cdelta'}.

\begin{IEEEproof}[Proof of the converse for Theorem~\ref{t_capacity}]
Fix an encoding scheme  $\{f_t\}_{t=0}^T$ with rate $R$ and probability of decoding error,
$P_{e}^{(T)} = P\{M \ne \hat{M} (Y_0^T)\}$. In addition, fix a partition $\tv$ of length $n$
such that $t_{i}-t_{i-1}<\Delta$  for any $i\in[1,2,\ldots,n]$ and let $t_n=T$. Consider
{\allowdisplaybreaks
\begin{align}
RT
&= H(M) \label{eq: converse a} \\
&= H(M)+H(M|Y_0^T)-H(M|Y_0^T) \\
&\le I(M;Y_0^T)+T\epsilon_T \label{eq: converse b} \\
&= I(M;Y_0^{t_1},Y_{t_1}^{t_2},\ldots,Y_{t_{n-1}}^{t_n})+T\epsilon_T \label{e_con1}
\\
&= \sum_{i=1}^n I(M;Y_{t_{i-1}}^{t_{i}}|Y_0^{t_{i-1}})+T\epsilon_T \\
&= \sum_{i=1}^n I(M,X_0^{t_{i-1}+\Delta};Y_{t_{i-1}}^{t_{i}}|Y_0^{t_{i-1}})+T\epsilon_T
   \label{eq: converse c} \\
&= \sum_{i=1}^n I(M,X_0^{t_i},X_{t_i}^{t_{i-1}+\Delta};Y_{t_{i-1}}^{t_{i}}|Y_0^{t_{i-1}})+T\epsilon_T \label{eq: converse d} \\
&= \sum_{i=1}^n I(M,X_0^{t_i};Y_{t_{i-1}}^{t_{i}}|Y_0^{t_{i-1}})+I(X_{t_i}^{t_{i-1}+\Delta};Y_{t_{i-1}}^{t_{i}}|Y_0^{t_{i-1}},M,X_0^{t_i})+T\epsilon_T \\
&= \sum_{i=1}^n I(X_0^{t_i};Y_{t_{i-1}}^{t_{i}}|Y_0^{t_{i-1}})+I(X_{t_i}^{t_{i-1}+\Delta};Y_{t_{i-1}}^{t_{i}}|Y_0^{t_{i-1}},M,X_0^{t_i})+T\epsilon_T \label{eq: converse e} \\
&= \sum_{i=1}^n I(X_0^{t_i};Y_{t_{i-1}}^{t_{i}}|Y_0^{t_{i-1}})+T\epsilon_T
   \label{eq: converse f} \\
&= I_\tv(X_0^T \to Y_0^T)+T\epsilon_T,\label{e_con2}
\end{align}}%
where the equality in~\eqref{eq: converse a} follows since the message is distributed uniformly,
the inequality in~\eqref{eq: converse b} follows from Fano's inequality, where
$\epsilon_T=\frac{1}{T}+P_e^{(T)}R$, the equality in~\eqref{eq: converse c} follows from the fact
that $X_0^{t_{i-1}+\Delta}$ is a deterministic function of $M$ and $Y_0^{t_{i-1}}$, the equality
in~\eqref{eq: converse d} follows from the assumption that $t_{i}-t_{i-1}<\Delta$, the equality
in~\eqref{eq: converse e} follows from \eqref{e_cdfy_Markov}, and the equality in~\eqref{eq: converse f}
follows from \eqref{e_cdfx_Markov}. Hence, we obtained that for every $\tv$
\begin{equation}
R \le \frac{1}{T}I_\tv(X_0^T \to Y_0^T)+\epsilon_T.
\end{equation}
Since the number of codewords is finite, we may consider the input signal of the form
$x_0^{T,\tv}$ with $x_{t_{i-1}}^{t_{i}}=f(u_0^T,y_0^{t_i-\Delta})$, where  the cardinality
of $u_0^T$ is bounded, i.e., $|\mathcal U_0^T| < \infty$ for any given $T$ (the bound may depend on $T$), independently of the
partition $\tv$. Furthermore,
\begin{align}
R&\le \inf_\tv \frac{1}{T}I_\tv( X_0^T \to Y_0^T)+\epsilon_T,\nonumber \\
&=\frac{1}{T}I( X_0^T \to Y_0^T)+\epsilon_T.
\end{align}
Finally, for any $R$ that is achievable there exists a sequence of codes such that
$\lim_{T\to\infty} P_e^{(T)}=0$, hence $\epsilon_T\to 0$ and we have established
\eqref{e_cdelta'}.
\end{IEEEproof}
Note that as a byproduct of the sequence of equalities \eqref{eq: converse b}--\eqref{e_con2}, we
conclude that for the
communication system depicted in Fig.~\ref{f_channel_con},
\begin{equation}\label{e_con3}
I(M;Y_0^T) = \inf_{\tv: t_i-t_{i-1}\le \delta} I_\tv(X_0^T \to Y_0^T)= I(X_0^T \to
Y_0^T).
\end{equation}
The only assumptions that we used to prove \eqref{eq: converse b}--\eqref{e_con2} is that the
encoders uses a strictly causal feedback of the form given in \eqref{e_xt_form} and that the
channel satisfies the benign assumption given in \eqref{e_cdfy_Markov}.  This might be a valuable
result by itself that provides a good intuition why directed information characterizes the
capacity of a continuous-time channel. Furthermore, the interpretations of the  measure
$I(M;Y_0^T)$, for instance, as given in \cite{Kadota--Zakai--Ziv1971b}, should also hold for
directed information and vice versa.


For the proof of achievability we will use the following result for discrete-time channels.

\begin{lemma}\label{l_ach}
Consider the discrete-time channel, where the input $U_i$ at time $i$ has a finite alphabet,
i.e.,  $|\mathcal U|<\infty$, and the output $Y_i$ at time $i$ has an arbitrary alphabet
$\mathcal Y$. We assume that the relation between the input and the output is given by
 \begin{equation} \label{disc-channel}
 Y_i=g(U_{i},Z_i),
 \end{equation}
 where the noise process $\{Z_i\}_{i\ge 1}$ is stationary and ergodic with an arbitrary alphabet $\mathcal Z$.
Then, any rate $R$ is achievable for this channel if
\begin{equation} \label{disc-rate}
R < \max_{p(u)} I(U;Y),
\end{equation}
where the joint distribution of $(U,Y)$ is induced by the input distribution $p(u)$, the stationary distribution of $Z$, and \eqref{disc-channel}.
\end{lemma}

\begin{IEEEproof}
Fix the pmf $p(u)$ that attains the maximum in \eqref{disc-rate}. Since $I(U;Y)$ can be
approximated arbitrarily close by a finite partition of $Y$ \cite{Gallager68}, assume without
loss of generality that $\mathcal{Y}$ is finite. The proof uses the random codebook generation
and joint typicality decoding in \cite[Ch.~3]{ElGammalKim10LectureNotes}. Randomly and
independently generate $2^{nR}$ codewords $u^n(m)$, $m = 1,2,\ldots, 2^{nR}$, each according to
$\prod_{i=1}^n p_U(u_i)$. The decoder finds the unique $\hat m$ such that $(u^n(m), y^n)$ is
jointly typical. (For the definition and properties of joint typicality, refer to \cite{Orlitsky--Roche2001}, \cite[Ch.~2]{ElGammalKim10LectureNotes}.) Now, assuming that $M = 1$ is sent, the decoder makes an error
only if $(U^n(1), Y^n)$ is not typical or $(U^n(m),Y^n)$ is typical for some $m \ne 1$. By the
packing lemma (\cite[Ch.~3]{ElGammalKim10LectureNotes}), the probability of the second event
tends to zero as $n \to \infty$ if $R < I(U;Y)$. To bound the probability of the first event,
recall from \cite[Th.~10.3.1]{Wolfowitz64} that if $\{U_i\}$ is i.i.d.\@ and $\{Z_i\}$ is
stationary ergodic, independent of $\{U_i\}$, then the pair $\{(U_i, Z_i)\}$ is jointly
stationary ergodic. Consequently, from the definition of the channel in \eqref{disc-channel},
$\{(U_i,Y_i)\}$ is jointly stationary ergodic. Thus, by Birkhoff's ergodic theorem, the
probability that $(U^n(1),Y^n)$ is not typical tends to zero as $n \to \infty$. Therefore, any
rate $R < I(U;Y)$ is achievable.
\end{IEEEproof}

The proof of achievability is based on the lemma above and the definition of directed information
for continuous time. It is essential to divide into small time-interval as well as
increasing the feedback delay by a small but positive value $\delta>0$.

\begin{IEEEproof}[Proof of achivability for Theorem \ref{t_capacity}]
Let $\Delta'=\Delta+\delta$, where $\delta>0$. In addition, let $\tv=(0 = t_0, t_1,\ldots,t_n = T)$ be
such that $t_i-t_{i-1}\le \delta$ for all $i=1,2,\ldots,n$. Let $X_0^{T,\tv}$ be of the form
\begin{equation}
X_{t_{i-1}}^{t_{i}} =
\begin{cases}
f(U_0^T,Y_0^{t_i-\Delta'})& t_i\ge \Delta',\\
f(U_0^T) & t_i<\Delta',
\end{cases}
\end{equation}
where the cardinality of $U_0^T$ is bounded. Then we show that
any rate
\begin{equation}\label{e_ach}
R < \frac{1}{T}I_\tv(X_0^{T,\tv} \to Y_0^T),
\end{equation}
is achievable.

Assume that the communication is over the time interval $[0,nT]$, where $T$ is fixed and $n$ may
be chosen to be as large as needed. Partition the time interval $[0,nT]$ into $n$ subintervals of
length $T$ and in each subinterval $[jT, jT+T)$, which we index by $j$, fix the relation
\begin{equation}
X_{jT+t_{i-1}}^{jT+t_{i}} =
\begin{cases}
f(U_{jT}^{jT+T},Y_{jT}^{jT+t_i-\Delta'})& t_i\ge \Delta',\\
f(U_{jT}^{jT+T}) & t_i<\Delta'.
\end{cases}
\end{equation}
Note that this coding scheme is possible with feedback delay $\Delta$ since $t_{i-1}-\Delta\ge
t_{i}-\Delta'$. This follows from the assumption that $t_i-t_{i-1}\le \delta$ and $\Delta'-\Delta
\ge \delta$. Now, let us define a discrete-time channel where the input at time $j+1$ is $\tilde
U_{j+1}=U_{jT}^{jT+T}$ (which has an alphabet $[1,\ldots,2^{nT}]$), the output at time $j+1$  is the
vector $\tilde Y_{j+1}=( Y_{jT}^{jT+t_1}, \ldots,Y_{jT+t_{i-1}}^{jT+t_i},\ldots,Y_{jT+t_{n-1}}^{jT+T})$
and the noise at time $j+1$ is $\tilde Z_{j+1}=Z_{jT}^{jT+T}$. Note that since $Z_{jT}^{jT+T}$ is
a stationary and block-ergodic  the noise process $\{\tilde Z_{j+1}\}_{j\ge0}$ is stationary and
ergodic. Furthermore the relation $\tilde Y_{j+1}=\tilde f(\tilde U_{j+1},\tilde Z_{j+1})$ holds
and the alphabet of $\tilde U_{j+1}$ is finite. Hence by Lemma \ref{l_ach}, any rate
\begin{equation}
R < \max_{p(\tilde u)} I(\tilde U;\tilde Y),
\end{equation}
is achievable. Now using the definition of the discrete-time channel and the properties of
directed information, we obtain
\begin{align}
I(\tilde U;\tilde Y)
&= I(U_0^T;Y_0^T) \label{eq ach proof step a} \\
&= I(U_0^T;Y_0^{t_1},Y_{t_1}^{t_2}, \ldots,Y_{t_n-1}^{t_n}) \\
&= I_\tv(X_0^{T,\tv} \to Y_0^{T,\tv}), \label{eq ach proof step b}
\end{align}
where the equality in \eqref{eq ach proof step a} follows from the definition of the
discrete-time channel and the equality in \eqref{eq ach proof step b} follows from the same
sequence of equalities as in \eqref{eq: converse b}--\eqref{e_con2}. Since \eqref{eq ach proof step b}
holds for any $\tv$ such that $t_i-t_{i-1}\le \delta$ we conclude that
\begin{equation} \label{e_ach3}
C(\Delta) \ge \inf_\tv I_\tv(X_0^T \to Y_0^T).
\end{equation}
Finally, by the definition of directed information and by the fact that \eqref{e_ach3} holds for
any $T$ we have established \eqref{e_cdelta'}.
\end{IEEEproof}

\section{Concluding Remarks}
\label{sec: Concluding Remarks}

We have introduced and developed a notion of directed information between
continuous-time stochastic processes. It emerges naturally in the  characterization
of the fundamental limit on reliable communication for a wide class of continuous-time channels with
feedback, quite analogously to the discrete-time setting. It also arises in estimation  theoretic relations as the replacement for mutual information when extending the scope to the presence of feedback. In particular, with
continuous-time directed information replacing mutual information, Duncan's theorem generalizes to estimation
problems in which the evolution of the target signal is affected by the past channel noise. An
analogous relationship based on the directed information holds for the Poisson channel.
We have illustrated the use of the latter in an explicit computation of the directed information rate between the input and output of
a Poisson channel  where the  input intensity changes only when there is an
event at the channel output.  One important direction for future exploration is to use the ``multiletter'' characterization of capacity developed here to compute or approximate the feedback capacity of interesting continuous-time channels.

\section*{Acknowledgments}
The authors thank the Associate Editor and the anonymous reviewers for their careful reading of the original manuscript and many valuable comments that helped improve the presentation.

\bibliographystyle{IEEEtran}

\end{document}